\documentclass[aps,prd,notitlepage,twocolumn,superscriptaddress,nofootinbib]{revtex4-1}
\usepackage[utf8]{inputenc}
\usepackage[english]{babel}
\usepackage{amsmath}
\usepackage{amsfonts}
\usepackage{amssymb}
\usepackage{listings}
\usepackage{lipsum}
\usepackage{multirow}
\usepackage{datetime}
\usepackage{graphicx}
\usepackage{mathtools}
\usepackage{mathrsfs}
\usepackage{dcolumn}
\usepackage{multirow}
\usepackage{color}   
\usepackage{hyperref}
\hypersetup{
    colorlinks=true, 
    pdfborder = {0 0 0.5 [3 3]},
    anchorcolor=black,
    citecolor=blue,
    linktoc=all,    
    linktocpage=true,
    linkcolor=red,
	urlcolor=blue
}

\newcommand{\figname}[1]{{{Fig.~#1}}}

\begin{document}

\title{Generic initial data for binary boson stars}
\author{Nils Siemonsen}
\email[]{nsiemonsen@perimeterinstitute.ca}
\affiliation{Perimeter Institute for Theoretical Physics, Waterloo, Ontario N2L 2Y5, Canada}
\affiliation{Arthur B. McDonald Canadian Astroparticle Physics Research Institute, 64 Bader Lane, Queen's University, Kingston, ON K7L 3N6, Canada}
\affiliation{Department of Physics \& Astronomy, University of Waterloo, Waterloo, ON N2L 3G1, Canada}
\author{William E.\ East}
\affiliation{Perimeter Institute for Theoretical Physics, Waterloo, Ontario N2L 2Y5, Canada}

\date{\today}

\begin{abstract} 
Binary boson stars can be used to model the nonlinear dynamics and
gravitational wave signals of merging ultracompact, but horizonless,
objects. However, doing so requires initial data satisfying the Hamiltonian
and momentum constraints of the Einstein equations, something that has not
yet been addressed. In this work, we construct constraint-satisfying
initial data for a variety of binary boson star configurations. We do this using the
conformal thin-sandwich formulation of the constraint equations, together
with a specific choice for the matter terms appropriate for scalar fields.
The free data is chosen based upon a superposition of isolated boson star
solutions, but with several modifications designed to suppress the spurious
oscillations in the stars that such an approach can lead to. We show that
the standard approach to reducing orbital eccentricity can be applied to
construct quasi-circular binary boson star initial data, reducing the
eccentricity of selected binaries to the $\sim 10^{-3}$ level.  Using these
methods, we construct initial data for quasi-circular binaries with
different mass-ratios and spins, including a configuration where the spin
is misaligned with the orbital angular momentum, and where the
dimensionless spins of the boson stars exceeds the Kerr bound. We evolve
these to produce the first such inspiral-merger-ringdown gravitational
waveforms for constraint-satisfying binary boson stars. Finally, we comment
on how equilibrium equations for the scalar matter could be used to improve
the construction of binary initial data, analogous to the approach used for
quasi-equilibrium binary neutron stars. 
\end{abstract}

\maketitle

\section{Introduction} \label{sec:Intro}

In recent years, the detection of gravitational wave (GW) signals have transformed the way we
observe the universe and probe fundamental physics. Predictions for the gravitational waveforms
of inspiraling and merging binary compact objects, such as black holes and
neutron stars, play a crucial role in the success of GW observation campaigns.
While during the early inspiral one can use perturbative methods, such an approach
breaks down as nonlinear effects take over at small separations of the
binary. This necessitates the use of numerical relativity to
accurately predict the merger dynamics and gravitational waveform through
merger. Crucially, numerical evolutions of binary compact objects rely on
initial data that satisfies the constraints of the
Einstein equations as a starting point. 
For binary black holes, binary neutron stars, and black hole-neutron star binaries,
there has been extensive work addressing the problem of constructing consistent initial
data (see, e.g.,
Refs.~\cite{Cook:2000vr,Pfeiffer:2004nc,Gourgoulhon:2007ue,Baumgarte:2010ndz}),
and a whole suite of
formalisms, numerical methods, and tricks have been developed (see
Ref.~\cite{Tichy:2016vmv} for a recent review) to obtain initial data suitable
for binary evolutions that can directly be compared to perturbative methods,
both in the late inspiral, and the ringdown phases of the coalescences (see,
e.g., Refs.~\cite{Pan:2013rra,Hannam:2013oca,Bohe:2016gbl}).
Thus, binary initial data is a crucial part of making the predictions that enable 
detecting GWs, characterizing GW sources, and
drawing conclusions for astrophysics and fundamental physics.

All GW observations of compact binary coalescence thus far are consistent with
arising from black hole and neutron stars.  However, besides these two classes,
(ultra) compact and black hole mimicking objects motivated by extensions of the
Standard Model and models of quantum gravity have been proposed to solve
various problems in high energy and particle physics \cite{Cardoso:2019rvt}.
Boson stars (BSs) are particularly simple examples of such ultracompact
objects
\cite{Kaup:1968zz,Ruffini:1969qy,Yoshida:1997qf,Seidel:1990jh,Seidel:1991zh,Brito:2015pxa},
and exhibit many common features of black hole mimickers, such as ergoregions,
stable light rings, and the absence of horizons
\cite{Friedberg:1986tq,Boskovic:2021nfs,Palenzuela:2017kcg,Kleihaus:2005me,Kleihaus:2007vk}
(see Refs.~\cite{Schunck:2003kk,Visinelli:2021uve} for reviews). Therefore,
considerable effort has been put into the development of a program to use
binary BSs as a simple test bed to study the nonlinear and highly dynamical
regimes of the larger class of exotic compact objects (see
Ref.~\cite{Liebling:2012fv} for a review). Progress has been made in
understanding the early inspiral of binary BSs and the resulting GW emission
using perturbative
methods~\cite{Macedo:2013jja,Sennett:2017etc,Krishnendu:2017shb,Johnson-Mcdaniel:2018cdu,Vaglio:2022flq,Adam:2022nlq,Pacilio:2020jza,Vaglio:2023lrd}.
There is also a significant body of work considering numerical relativity
evolutions of scalar and vector binary BSs, ranging from head-on collisions of
non-spinning and spinning binaries
\cite{Lai:2004fw,Choptuik:2009ww,Paredes:2015wga,Bernal:2006ci,Schwabe:2016rze,Palenzuela:2006wp,Mundim:2010hi,Bezares:2017mzk,Sanchis-Gual:2018oui,Helfer:2021brt,Sanchis-Gual:2022mkk,Evstafyeva:2022bpr},
to orbital inspirals
\cite{Palenzuela:2007dm,Palenzuela:2017kcg,Bezares:2022obu,Bezares:2018qwa,Bezares:2017mzk,Sanchis-Gual:2018oui,Siemonsen:2023hko}.
However, despite the considerable amount of work on the binary BS evolution
problem, the problem of constructing initial data for these compact binaries
that is consistent with the Einstein equations has yet to be addressed. 

The common practice in the literature, when constructing binary BS initial data,
has been so far to simply superpose two boosted star solutions.  More recently, in
Refs.~\cite{Helfer:2021brt,Evstafyeva:2022bpr}, a modified superposition trick
was utilized to reduce (but not eliminate) violations of the Hamiltonian and
momentum constraints.  While simple, superposed initial data leads to large 
constraint violations even at moderate separations, meaning 
that solutions to the evolution equations do not accurately approximate solutions
of the Einstein equations and effectively excluding, for 
example, the quasi-circular binaries relevant to current GW observations of
inspiral-merger-ringdown\footnote{Note that this can not necessarily be addressed by including
constraint damping terms in the evolution equations. See Appendix~\ref{app:numerics}.
}.  

In this work, we develop and implement methods for
constructing constraint-satisfying binary scalar BS initial data for a wide
variety of configurations. (We also used these methods recently in
Ref.~\cite{Siemonsen:2023hko}.) We solve the constraint equations in the
conformal thin-sandwich (CTS) formalism~\cite{York:1998hy}, using free data
based on superposing stationary BS solutions, similar to what was done in
Ref.~\cite{East:2012zn} for black hole and fluid stars. However, considering
scalar matter introduces several new complications which we address here,
including the choice of which matter degrees of freedom to fix, as well as how
to minimize spurious oscillations which may be induced in the BSs.  Our
approach is very flexible, and we use it to construct binary BSs with different
mass-ratios, spin magnitudes, and spin orientations.  We evolve several such
binaries, including several cases where the BSs are super-spinning (i.e., have
dimensionless spins exceed the Kerr bound of unity), through inspiral and
merger. We do this both for physical interest, as well as to demonstrate that
we can construct quasi-circular binaries by adapting eccentricity reduction
techniques.

The remainder of this paper is organized as follows.  We briefly review the relevant
physics of isolated stationary BSs in Sec.~\ref{sec:isolatedstar}, and proceed
in Sec.~\ref{sec:ctsformulation} to introduce our procedure to
self-consistently solve the elliptic constraint equations in the CTS
formalism~\cite{York:1998hy} numerically, given an initial guess for the
binary, utilizing the elliptic solver introduced in Ref.~\cite{East:2012zn}. To
that end, we identify the most suitable parameterization of the matter content
of BSs in Sec.~\ref{sec:bbssource}. We comment on possible equilibrium
conditions for the scalar matter of these stars in
Sec.~\ref{sec:scalarequilibrium}, though we do not implement such an approach
in this study.  In Sec.~\ref{sec:spurious_osc}, we analyze the quality of the
constructed initial data and devise methods to reduce spurious oscillations in
each star, as well as in the resulting gravitational radiation, and lastly, in
Sec.~\ref{sec:ecc_red}, we test eccentricity reduction schemes in the context
of binary BS inspirals and comment on their possible limitations.  We consider
binary configurations in two different scalar potentials, with equal and
unequal masses, as well as non-spinning and spinning constituent stars with
aligned and misaligned spins. Finally, in Sec.~\ref{sec:binaryevolutions}, we
analyze the dynamics of selected eccentricity-reduced binary configurations, and
present inspiral-merger-ringdown gravitational waveforms.  We give details on
the numerical evolution scheme, present convergence results, and briefly compare 
the constraint satisfying initial data we construct in this study to superposed initial data in 
Appendix~\ref{app:numerics}. We provide some further details on spurious high
frequency components to the GWs and correcting for center-of-mass motion of the
BS binaries in Appendices~\ref{app:contamination}
and~\ref{app:centerofmassmotion}, respectively.  We use $G=c=1$ units
throughout.

\section{Methodology} \label{sec:method}

\subsection{Isolated boson star} \label{sec:isolatedstar}

Before turning to the construction of binary BS data, we briefly review isolated BSs in their rest frames. Here and throughout, we focus entirely on scalar BSs, and leave a full consideration of vector BSs to future work. Scalar BSs are regular, asymptotically flat, stationary, and axisymmetric solutions to the globally U(1)-invariant Einstein-Klein-Gordon theory
\begin{align}
S=\int d^4x \sqrt{-g}\left[ \frac{R}{16\pi}-g^{\alpha\beta}\nabla_{(\alpha}\bar{\Phi}\nabla_{\beta)}\Phi-V(|\Phi|)\right],
\label{eq:action}
\end{align}
where $R$ is the Ricci scalar of the spacetime $g_{\mu\nu}$, $\Phi$ is the complex scalar field (with an overbar denoting complex conjugation), and $V(|\Phi|)$ is the global U(1)-preserving scalar potential. 
We consider cases where the lowest order term in $|\Phi|$ of $V$ is a mass term $\mu^2 |\Phi|^2$. 
The conserved Noether current associated with the $U(1)$ symmetry of the action~\eqref{eq:action} is
\begin{align}
\nabla_\mu j^\mu=0, & & j^\mu=-i(\bar{\Phi}\nabla^\mu\Phi-\Phi\nabla^\mu\bar{\Phi}).
\label{eq:Noetherconservation}
\end{align}
Intuitively, the charge $Q$ of a BS counts the number bosons in the solution. This is in direct analogy to the conservation of baryon number in the case of fluid stars. 

The metric ansatz for isolated star solutions in Lewis-Papapetrou coordinates takes the form
\begin{align}
\begin{aligned}
ds^2=-fdt^2+lf^{-1}\big\{ g(dr^2+r^2d\theta^2)\\
+r^2\sin^2\theta(d\varphi-\Omega r^{-1} dt)^2\big\}
,
\label{eq:metricansatz}
\end{aligned}
\end{align}
where the metric functions $f$, $l$, $g$, and $\Omega$ depend, in general, on both $r$ and $\theta$. The boundary conditions on these functions for BS solutions can be found in Ref.~\cite{Kleihaus:2005me}; generally, the solution is regular at the star's center and asymptotically flat. The scalar field ansatz for BSs is
\begin{align}
\Phi=\phi(r,\theta) e^{i(\omega t-m\varphi)},
\label{eq:scalaransatz}
\end{align}
where $\phi$ is the magnitude, while $\omega$ is the star's internal frequency, and $m$ is the integer azimuthal index. At large distances, due to the non-zero scalar mass, the field falls off exponentially as
\begin{align}
\lim_{r\rightarrow\infty}\phi\sim e^{-\sqrt{\mu^2-\omega^2}r},
\end{align}
ensuring that the solution is asymptotically flat. Spherically symmetric star solutions attain a non-zero scalar field magnitude at the center, $\phi|_{r=0}\neq 0$, while rotating solutions (i.e., those with $|m|\geq 1$) exhibit a vortex line through their centers, and hence, are toroidal in shape (see Ref.~\cite{Siemonsen:2023hko} for a discussion of the vortex structure). Details on the numerical construction of these solutions used in this work are discussed in the appendices of Ref.~\cite{Siemonsen:2020hcg}.

We are interested in testing our methods to construct binary BS initial data in various physically interesting regimes. Therefore, we focus entirely on scalar models with non-vanishing self-interactions, as these have shown to lead to highly-compact \cite{Palenzuela:2017kcg,Boskovic:2021nfs}, as well as stably rotating BS solutions \cite{Siemonsen:2020hcg}. To that end, we consider the \textit{solitonic} scalar potential \cite{Friedberg:1986tq}
\begin{align}
V(|\Phi|)=\mu^2|\Phi|^2\left(1-\frac{2|\Phi|^2}{\sigma^2}\right)^2,
\label{eq:solitonicpotential}
\end{align}
controlled by the coupling $\sigma$, as well as the leading repulsive scalar self-interaction
\begin{align}
V(|\Phi|)=\mu^2|\Phi|^2+\lambda|\Phi|^4,
\label{eq:repulsivepotential}
\end{align}
in the following simply referred to as the \textit{repulsive} potential
(restricted to $\lambda>0$). Imposing the appropriate asymptotically flat
condition at spatial infinity, and regularity conditions at the origin, in
conjunction with the ansätze \eqref{eq:metricansatz} and \eqref{eq:scalaransatz},
and the field equations resulting from \eqref{eq:action}, yields a
one-parameter family of solutions for each azimuthal index $m$ parameterized by
the internal frequency $\omega$\footnote{We focus solely on stars in their
respective radial ground states, i.e., the scalar field profiles have no
additional radial nodes.}. For each member of the family of spacetimes, the
total mass $M$ is given by the Komar expression, coinciding with the ADM mass,
and the radius $R$ is defined as the areal (circular) radius $r_*$ at which $99\%$
of the mass of the spherical (rotating) star solution lies within $r_*$. 
The angular momentum $J$ is well-defined by means of the
Komar expression in this stationary and axisymmetric class of spacetimes.
Finally, the charge $Q$ of each star follows from the U(1)-Noether charge of
the action \eqref{eq:action} (details can be found in
Ref.~\cite{Siemonsen:2020hcg}). For these isolated BS solutions, this Noether
charge satisfies the ``quantization" relation $J=mQ$.

\begin{figure}[t]
\centering
\includegraphics[width=0.485\textwidth]{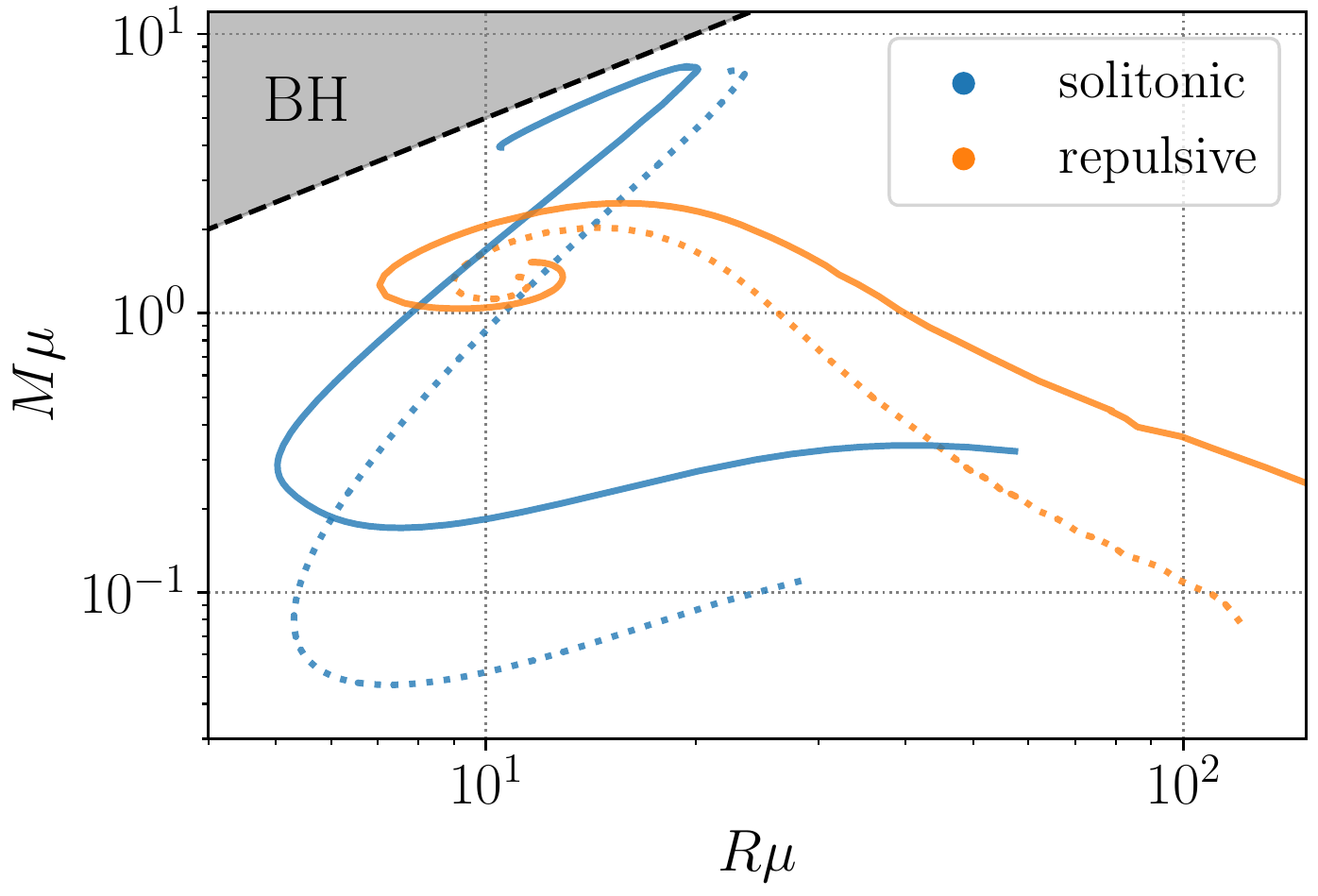}
\caption{We plot the relationship between the mass $M$ and the radius $R$, in units of the boson mass parameter $\mu$, for four families of BS solutions. We show the solitonic scalar model \eqref{eq:solitonicpotential} with coupling $\sigma=0.05$ \textit{(blue curves)}, and the repulsive potential \eqref{eq:repulsivepotential} with coupling $\lambda/\mu^2=10^3$ \textit{(orange curves)}, including both spherically symmetric \textit{(dotted)} and $m=1$ rotating families of stars \textit{(solid)}. The regime with $M/R>1/2$ is labelled as ``BH."}
\label{fig:rvmplot}
\end{figure}

In \figname{\ref{fig:rvmplot}}, we show the mass-radius relationships of the isolated BS solutions considered in this work. As can be seen there, there is a regime where the solitonic families have highly compact solutions; as a result, some of these solutions exhibit regions of stable trapping of null geodesics, as well as large ergoregions if $m>0$. Hence, this class of spacetimes exhibit highly relativistic features potentially leading to interesting phenomenology in the context of binary systems. In contrast to this, solutions of families in the scalar model with repulsive self-interactions are mostly less compact, $C_{\rm max}\approx 0.14$, enabling the study of BSs in the Newtonian regime. Importantly, both scalar models contain spinning solutions with $m=1$ that have been shown to be stable \cite{Siemonsen:2020hcg} against the non-axisymmetric instability discovered in Ref.~\cite{Sanchis-Gual:2019ljs} on long timescales.

\subsection{Conformal thin-sandwich formulation} \label{sec:ctsformulation}

Moving now to the formalism used to construct constraint satisfying binary BS initial data, we introduce the CTS formulation of the Hamiltonian and momentum constraints of the Einstein equations. To that end, the spacetime is foliated into a series of spacelike hypersurfaces $\Sigma_t$, parameterized by the coordinate time $t$, with future-pointing unit-normal to the hypersurface $n^\mu$. The tangent of lines of constant spatial coordinates is then
\begin{align}
t^\mu=\alpha n^\mu+\beta^\mu,
\label{eq:timevector}
\end{align}
with lapse function $\alpha$ and shift vector $\beta^\mu$, with $n_\mu \beta^\mu=0$. Furthermore, let $\gamma_{\mu\nu}=g_{\mu\nu}+n_\mu n_\nu$ be the projector onto the hypersurface $\Sigma_t$, such that $\gamma_{ij}$ is the spatial metric induced on $\Sigma_t$. Lastly, the extrinsic curvature of $\Sigma_t$,
\begin{align}
K_{ij}=-\frac{1}{2}\mathcal{L}_n\gamma_{ij},
\end{align}
is defined by means of the Lie derivative $\mathcal{L}_n$ along the hypersurface normal. In this 3+1 language, the Hamiltonian and momentum constraints, i.e., the projections of the Einstein equations along the hypersurface normal, are
\begin{align}
\begin{aligned}
{}^{(3)}R+K^2+K_{ij}K^{ij}= & \ 16\pi E, \\
D_j K^{ij}-D^i K= & \ 8\pi p^i,
\label{eq:constraints}
\end{aligned}
\end{align}
with trace $K=\gamma^{ij}K_{ij}$, Ricci scalar ${}^{(3)}R$ and derivative $D_i$ defined with respect to the induced metric $\gamma_{ij}$, and finally, the energy density $E=n_\mu n_\nu T^{\mu\nu}$ and momentum density $p^i=-\gamma^i_\mu n_\nu T^{\mu\nu}$ of the matter content of the space in the Eulerian frame.

The CTS formulation \cite{York:1998hy} of the Hamiltonian and momentum constraints \eqref{eq:constraints} relies on relating the constraint satisfying metric components $\gamma_{ij}$ to a freely specifiable conformal metric $\tilde{\gamma}_{ij}$ as
\begin{align}
\gamma_{ij}=\Psi^4\tilde{\gamma}_{ij},
\end{align}
with conformal factor $\Psi$. Furthermore, the traceless part of the extrinsic curvature $A^{ij}$ is conformally decomposed as $A^{ij}=\Psi^{-10}\hat{A}^{ij}$ with
\begin{align}
\hat{A}^{ij}=\frac{1}{2\tilde{\alpha}}\Big[(\tilde{\mathbb{L}}\beta)^{ij}+\partial_t\tilde{\gamma}^{ij}\Big],
\end{align}
in terms of the conformal Killing form $(\tilde{\mathbb{L}}\beta)^{ij}=\tilde{D}^i\beta^j+\tilde{D}^j\beta^i-2\tilde{\gamma}^{ij}\tilde{D}_k\beta^k/3$. The conformal lapse and the time-derivative of the conformal metric are $\tilde{\alpha}=\Psi^{-6}\alpha$ and $\partial_t\tilde{\gamma}^{ij}=\Psi^4(\partial_t\gamma^{ij}-\gamma^{ij}\gamma_{kl}\partial_t\gamma^{kl}/3)$, respectively. Utilizing this decomposition, the metric constraints \eqref{eq:constraints} are cast into the CTS equations
\begin{align}
\begin{aligned}
\tilde{D}_i\tilde{D}^i\Psi-\frac{\tilde{R}}{8}\Psi+\frac{\hat{A}_{ij}\hat{A}^{ij}}{8}\Psi^{-7}-\frac{K^2}{12}\Psi^5= & -2\pi\Psi^{5}E,\\
\tilde{D}_j\hat{A}^{ij}-\frac{2}{3}\Psi^6\tilde{D}^iK= & \ 8\pi\Psi^{10}p^i.
\label{eq:ctsequations}
\end{aligned}
\end{align}
The geometric free data (i.e., those metric variables that need to be specified) are comprised of the metric $\tilde{\gamma}_{ij}$ and its coordinate time derivative $\partial_t\tilde{\gamma}^{ij}$, as well as the trace $K$ of the extrinsic curvature and the lapse function $\tilde{\alpha}$. This formulation is supplemented by a choice of energy and momentum densities, $E$ and $p^i$, of the complex scalar matter, as well as the corresponding scalar free data. This will be discussed in detail in the next section.

To specify the metric free data, we proceed as follows. First, we solve for
stationary isolated BSs in Lewis-Papapetrou coordinates, as outlined in
Sec.~\ref{sec:isolatedstar}. These solutions are subsequently transformed to
Cartesian coordinates\footnote{We transform from Lewis-Papapetrou to Cartesian
coordinates by applying the usual flat relations between spherical and
Cartesian coordinates.} and boosted using initial coordinate velocities
$v_{(A)}^i$, where $A\in\{1,2\}$ labels each star in the binary, and placed at
coordinate positions $z_{(A)}^i$. Therefore, for each star we obtain the set of
variables $\gamma_{ij}^{(A)}$, $\partial_t\gamma_{ij}^{(A)}$, $\alpha^{(A)}$,
and $\beta^i_{(A)}$ (for both stars in Cartesian-type coordinates). For all binary
configurations presented in this work, $z_{(A)}^i$ and $v_{(A)}^i$ are chosen
such that the initial center-of-mass location coincides with the origin of the
numerical grid, and the initial linear momentum of the center of mass vanishes
(at the Newtonian level); limitations of this approach are discussed in
Appendix~\ref{app:centerofmassmotion}. These two solutions are then superposed as
\begin{align}
\begin{aligned}
\gamma_{ij}^{\rm sup}= & \ \eta_{ij}+f_{(2)}\big[\gamma^{(1)}_{ij}-\eta_{ij}\big]+f_{(1)}\big[\gamma^{(2)}_{ij}-\eta_{ij}\big], \\
\partial_t\gamma^{\rm sup}_{ij}= & \ f_{(2)}\partial_t\gamma_{ij}^{(1)}+f_{(1)}\partial_t\gamma_{ij}^{(2)}, \\
\alpha_{\rm sup}= & \ 1+f_{(2)}\big[\alpha^{(1)}-1\big]+f_{(1)}\big[\alpha^{(2)}-1\big], \\
\beta^{i}_{\rm sup}= & \ f_{(2)}\beta^{i}_{(1)}+f_{(1)}\beta^{i}_{(2)},
\label{eq:superposemetricdata}
\end{aligned}
\end{align}
where $\eta^i{}_{j}=\delta^i_j$ is the flat 3-metric, and $f_{(A)}$ is an attenuation function, which we introduce here for convenience and discuss in detail in Sec.~\ref{sec:spurious_osc}. For now, we simply point out that the choice $f_{(A)}\equiv 1$ corresponds to a plain superposition of the isolated stars. As discussed in Ref.~\cite{East:2012zn}, the metric free data is then obtained from
\begin{align}
\begin{aligned}
\tilde{\gamma}_{ij}= & \ \gamma_{ij}^{\rm sup}, \\
    \partial_t\tilde{\gamma}^{ij}=  & -\tilde{\gamma}^{ik}\tilde{\gamma}^{jl}\left(\partial_t\gamma_{kl}^{\rm sup}-\frac{1}{3}\tilde{\gamma}_{kl}\tilde{\gamma}^{mn}\partial_t\gamma_{mn}^{\rm sup}\right),  \\
\tilde{\alpha}= & \ \alpha_{\rm sup}, \\
K= & \ \frac{1}{2\tilde{\alpha}}\big[2\partial_i\beta^i_{\rm sup}+\tilde{\gamma}_{ij}\partial_t\tilde{\gamma}^{ij}+\tilde{\gamma}^{ij}\beta^k_{\rm sup}\partial_k\tilde{\gamma}_{ij} \big].
\label{eq:metricfreedata}
\end{aligned}
\end{align}

The CTS equations admit solutions provided appropriate boundary conditions are specified. In the context of binary BSs, i.e., asymptotically flat spacetimes, we require that 
\begin{align}
\lim_{|\textbf{x}|\rightarrow\infty}\Psi=1, & & \lim_{|\textbf{x}|\rightarrow\infty}\beta^i=\beta^i_{\rm sup}|_\infty,
\end{align}
where $\beta^i_{\rm sup}|_\infty$ is the shift of the free data at large distances. With these boundary conditions, we solve the CTS equations numerically using a multigrid scheme with fixed mesh refinement (further details can be found in Ref.~\cite{East:2012zn}). In the context of axisymmetry, we employ a generalized Cartoon method that provides derivatives about the axis of symmetry by means of a the axisymmetric Killing field, allowing also for harmonic azimuthal dependencies in the scalar sector.

\subsection{Binary boson star sources} \label{sec:bbssource}

So far, we have left the precise parameterization of the scalar matter sourcing the spacetime, $E$ and $p^i$, unspecified. In principle, various choices of energy and momentum densities measured by an Eulerian observer, are possible for time-dependent complex scalar field matter. However, we find the precise choice to be crucial to achieve convergence of our numerical implementation. Therefore, in the following we outline possible matter source parameterizations, and especially, highlight the method we found to robustly yield consistent binary BS initial data in any considered context.

We begin by introducing the necessary projections of the nonlinear complex scalar energy-momentum tensor with respect to the foliation introduced in the previous section. The latter is readily obtained from \eqref{eq:action} in covariant form:
\begin{align}
T_{\mu\nu}= 2\partial_{(\mu}\bar{\Phi}\partial_{\nu)}\Phi -g_{\mu\nu}\Big[g^{\alpha\beta}\partial_{(\alpha}\bar{\Phi}\partial_{\beta)}\Phi+V(|\Phi|) \Big].
\label{eq:energymomentumtensor}
\end{align}
For convenience, we define the conjugate momentum of the complex field with respect to a spatial slice as
\begin{align}
\eta:=\mathcal{L}_n\Phi=n^\alpha\partial_\alpha\Phi=\frac{1}{\alpha}(\partial_t\Phi-\beta^iD_i\Phi).
\label{eq:kineticenergy}
\end{align}
With this, the scalar energy-momentum tensor can be written in the Eulerian frame as
\begin{align}
\begin{aligned}
E=& \ n_\alpha n_\beta T^{\alpha\beta}=\eta\bar{\eta}+D^i\Phi D_i\bar{\Phi}+V(|\Phi|)\\
p^i=& -\gamma^i{}_\alpha n_\beta T^{\alpha\beta}=-\eta D^i\bar{\Phi}-\bar{\eta}D^i\Phi\\
S^{ij}= & \ \gamma^i{}_\alpha \gamma^j{}_\beta T^{\alpha\beta} = 2D^{(i}\Phi D^{j)}\bar{\Phi}\\
& \ + \gamma^{ij}\Big[\eta\bar{\eta}-D_k\Phi D^k\bar{\Phi}-V(|\Phi|)\Big].
\label{eq:energymomentumdensities}
\end{aligned}
\end{align}
Starting from these expressions, we discuss possible approaches to parameterize the scalar matter, as well as the associated choices of scalar free data accompanying the metric free data \eqref{eq:metricfreedata}. To that end, and before presenting the source parameterization that we found to work for any type of binary BS configuration, it is instructive to consider two other approaches that, while natural, exhibit fundamental issues.

\subsubsection{Fixed energy and momentum densities} \label{sec:fixedenergymomentum}

In the context of binary neutron star initial data, it is natural to choose the
conformal Eulerian energy $\tilde{E}$ and momentum densities $\tilde{p}^i$ as
free data for the CTS system of equations \eqref{eq:ctsequations}. The
corresponding physical momentum density $p^i$ is typically chosen to be
$p^i=\Psi^{-10}\tilde{p}^i$.  One choice for the conformal scaling of the
energy is $E=\Psi^{-8}\tilde{E}$, which is motivated by uniqueness arguments
and the preservation of the dominant energy condition (i.e., if the free data
satisfies $\tilde{E}\geq \sqrt{\tilde{\gamma}^{ij}\tilde{p}_i\tilde{p}_j}$ so
does the constraint satisfying initial data). In the case of fluid stars, the
initial physical pressure $P$ and density $\rho$ are recovered by means of
an algebraic relation between $E$, $p^i$, $P$, and $\rho$ derived from the expression for the 
fluid energy-momentum tensor combined with the fluid equation of state. This provides a means to reconstruct the
constraint satisfying fluid variable initial conditions directly from the free data
and constrained data.

A complex scalar field, on the other hand, has kinetic and gradient energy, in addition to potential energy. The energy therefore depends on spatial gradients $D^i\Phi$ and time-derivatives $\partial_t\Phi$ of the scalar field, as can be seen in \eqref{eq:energymomentumdensities}. Therefore, unlike in the binary neutron star scenario, the relation between the physical energy and momentum densities, and the matter field $\Phi$, is not purely algebraic, but rather of differential form. This renders the reconstruction of the scalar field initial data $\Phi$ and  $\partial_t\Phi$ (or equivalently $\Phi$ and $\eta$) from the constraint satisfying energy and momentum densities non-trivial. 

Irrespective of these shortcomings, we test this choice of matter source variables with a set of single isolated non-spinning and spinning BSs. The metric free data is constructed following the discussion in Sec.~\ref{sec:ctsformulation} (setting $\gamma_{ij}^{(2)}=0$ etc.), while the scalar sources $\tilde{E}$ and $\tilde{p}^i$ are determined from \eqref{eq:energymomentumdensities}. The CTS equations are then numerically solved iteratively as outlined above. We succeeded in recovering isolated, boosted, non-spinning and spinning BSs utilizing this approach, i.e., the elliptic CTS solver removed truncation error of the isolated solution to the precision allowed by the resolution of the discretization of the CTS equations. This shows that, within our numerical setup, solutions to isolated stars are in fact local attractors in the space of solutions using this scalar matter parameterization.

\subsubsection{Fixed scalar initial data} \label{sec:fixedscalartimederivative}

In order to circumvent the issue discussed in the previous section, i.e., instead of fixing the energy and momentum densities directly, one could provide the scalar field initial data itself---$\Phi$ and $\partial_t\Phi$---as free data to the CTS equations. This makes the reconstruction of the scalar field trivial, ensuring that the metric and scalar initial data consistently solve the Hamiltonian and momentum constraints. To that end, we rewrite \eqref{eq:energymomentumdensities} in terms of the scalar fields $\Phi$ and $\partial_t\Phi$, leading to
\begin{align}
\begin{aligned}
E= & \ \frac{\Psi^{-12}}{\tilde{\alpha}^2}|\partial_t\Phi-\beta^i\partial_i\Phi|^2+\Psi^{-4}\tilde{D}^i\bar{\Phi}\tilde{D}_i\Phi+V(|\Phi|),\\
 p^i= & -\frac{\Psi^{-10}}{\tilde{\alpha}^2}\Big((\partial_t\Phi-\beta^i\partial_i\Phi)\tilde{D}^i\bar{\Phi}+(\partial_t\bar{\Phi}-\beta^i\partial_i\bar{\Phi})\tilde{D}^i\Phi\Big)
 \label{eq:scalarinitialdatasources}
\end{aligned}
\end{align}
in terms of the conformal variables. The different scaling of the kinetic, gradient, and potential energies with conformal factor $\Psi$, as well as the dependencies on the shift vector $\beta^i$ indicate that this approaches differs from providing $E$ and $p^i$ as free data by more than a simple $\Psi$-rescaling. 

We tested the above choice of sources within our numerical setup, similarly to
our tests of the formulation presented in Sec.~\ref{sec:fixedenergymomentum}.
We found robust convergence of the numerical schemes in the case of isolated
boosted non-spinning BSs. However, we were unable to recover an isolated
stationary rotating BS solution using \eqref{eq:scalarinitialdatasources} in
the CTS equations. Despite the CTS equation residuals converging to zero at the
expected order \textit{before} the first iteration, the solution of the
elliptic solver moves away from the true solution exponentially quickly with
each iteration. This indicates that rotating BSs are not attractors in the
space of solutions when using \eqref{eq:scalarinitialdatasources} and our
numerical framework, or suggests a break-down in the uniqueness of this solution
for the given free and boundary data. Note, tests of uniqueness based on the
maximum principle (see, e.g., Ref.~\cite{Gourgoulhon:2007ue}) are not
applicable in this case, since the momentum constraint is not trivially
satisfied by stationary rotating BS solutions\footnote{Even if the maximum
principle could be applied in this case, perturbations $\epsilon$ away from a
solution $\Psi_0$ to the Hamiltonian constraint follow to linear order the
equation $\tilde{D}_i\tilde{D}^i\epsilon=\Gamma\epsilon$, with $\Gamma=
\tilde{R}/2
-14\pi\Psi_0^{-8}\tilde{\alpha}^{-2}|\partial_t\Phi-\beta^i\partial_i\Phi|^2+\dots$,
where we ignored all positive-definite terms. Hence, non-vanishing kinetic
energy and momentum densities $p^i$, present even in isolated rotating BSs, may
result in violations of the maximum principle, i.e., in $\Gamma<0$.}. 

\subsubsection{Fixed scalar kinetic energy} \label{sec:fixedkineticenergy}

We turn now to the choice of scalar matter free data that we found to robustly lead to constraint satisfying binary BS metric and scalar field initial data. Similar choices of free data were considered recently in cosmological contexts in Refs.~\cite{Aurrekoetxea:2022mpw,Corman:2022alv}. Here, instead of setting the scalar initial data $\{\Phi,\partial_t\Phi\}$ as free data for the CTS equations, we replace $\partial_t\Phi$ by the scalar field's conformal conjugate momentum $\tilde{\eta}=\Psi^6 \eta$ as free data. With this, and in terms of the above conformal decomposition, the energy and momentum densities turn into
\begin{align}
\begin{aligned}
E= & \ \Psi^{-12}\tilde{\eta}\bar{\tilde{\eta}}+\Psi^{-4}\tilde{D}^i\Phi\tilde{D}_i\bar{\Phi}+V(|\Phi|), \\
p^i= & \ -\Psi^{-10}(\tilde{\eta}\tilde{D}^i\bar{\Phi}+\bar{\tilde{\eta}}\tilde{D}^i\Phi).
\label{eq:fixedkinenergy}
\end{aligned}
\end{align}
The scalar data satisfying the constraint equations can then be recovered via the \textit{algebraic} relation
\begin{align}
\partial_t\Phi=\tilde{\alpha}\tilde{\eta}+\beta^i\partial_i\Phi,
\label{eq:scalarfieldtimederivative}
\end{align}
where $\beta$ is the solution to the vector CTS equation provided the free data $\{\Phi,\tilde{\eta}\}$; notice $\alpha\eta=\tilde{\alpha}\tilde{\eta}$. For completeness, we include here also the expressions \eqref{eq:fixedkinenergy} in terms of $\Phi_R=(\Phi+\bar{\Phi})/2$, and $\Phi_I=(\Phi-\bar{\Phi})/(2i)$, as these are the variables used in our numerical implementation of the elliptic CTS solver, as well as the hyperbolic evolution scheme:
\begin{align}
\begin{aligned}
E= & \ \big[\Psi^{-12}\tilde{\eta}_R^2+\Psi^{-4}\tilde{D}_i\Phi_R\tilde{D}^i\Phi_R+(R\leftrightarrow I)\big]\\
& \qquad\qquad +V(\Phi_R^2+\Phi_I^2), \\
p^i=& -2\Psi^{-10}\big[\tilde{\eta}_R\tilde{D}^i \Phi_R +(R\leftrightarrow I)\big].
\end{aligned}
\end{align}
We found this matter source parameterization to robustly recover any kind of single BS solution in those tests outlined in Sec.~\ref{sec:fixedenergymomentum}. Given this parameterization passes these tests, we are now able to move to \textit{binary} BS. To that end, we construct the scalar free data in a similar fashion to the superposed metric free data presented in \eqref{eq:superposemetricdata}. First, the scalar fields $\Phi^{(A)}$ of each star are boosted with the same boost as the metric, then the conjugate momenta $\tilde{\eta}^{(A)}$ are determined\footnote{Recall, the scalar field has a non-trivial time-dependence [see also \eqref{eq:scalaransatz}], so that we first boost the vector $\partial_\mu \Phi$, and then determine its conjugate momentum in the boosted frame.}, and finally, the variables are superposed to obtain the scalar free data as follows:
\begin{align}
\begin{aligned}
\Phi^{\rm sup}= & \ \hat{f}_{(2)}\Phi^{(1)}+\hat{f}_{(1)}\Phi^{(2)}, \\
\tilde{\eta}^{\rm sup}= & \ \hat{f}_{(2)}\tilde{\eta}^{(1)}+\hat{f}_{(1)}\tilde{\eta}^{(2)}.
\label{eq:scalarsuperposition}
\end{aligned}
\end{align}
Here $\hat{f}_{(A)}$ are attenuation functions (directly analogous to $f_{(A)}$, defined in Sec.~\ref{sec:ctsformulation}), which we discuss in detail below and simply note here that $\hat{f}_{(A)}=1$ corresponds to a simple superposition of the two star's scalar field variables. Notice, while the constraint equations are invariant under a global phase shift $\Phi\rightarrow\Phi e^{i\alpha}$, the source functions \eqref{eq:fixedkinenergy} are not invariant under the phase of a \textit{single} constituent of a binary, e.g., $\Phi^{(1)}\rightarrow \Phi^{(1)}e^{i\alpha}$.

\subsection{Scalar matter equilibrium} \label{sec:scalarequilibrium}

We focused so far on finding a scalar matter parameterization that robustly
yields constraint satisfying binary BS initial data. Since no assumptions on
the stars' trajectories, spin orientations, or mass-ratio were built into the
formalism, this approach is very flexible. However, as we show in
Sec.~\ref{sec:qualityofinitialdata}, the simple construction of the free data
described above results in stars with large internal oscillations and ejected
scalar matter. In the case of binary neutron stars, various methods have been
introduced to alleviate these issues by explicitly equilibriating the fluid and
metric degrees of freedom
\cite{Bonazzola:1997gc,Shibata:1998um,Teukolsky:1998sh}. These 
approaches are based on assuming the existence of a helical Killing field $\ell^\mu$,
which provides a notion of equilibrium not just for the metric, i.e.,
$\mathcal{L}_\ell g_{\mu\nu}=0$, but also for the matter variables. In the case
of binary neutron stars, combining the conservation of rest mass density, the
conservation of the fluid's energy-momentum, and matter equilibrium with
respect to $\ell^\mu$, results in an elliptic equation for the equilibriated
initial velocity of the fluid. In the following, we apply these arguments
qualitatively to the case of scalar field matter and outline approaches to
equilibriating the scalar matter. However, as we are not testing these formalisms
here explicitly, this is to be understood as a first step guiding more thorough
future analyses.

To that end, we assume the existence of a helical Killing field $\ell^\mu$, such that $\ell^\mu=\alpha n^\mu+\tilde{V}^\mu$. In the asymptotically inertial frame, the spatial velocity takes the form $\tilde{V}^\mu=\beta^\mu+\tilde{\Omega}m^\mu$, where $m^\mu$ is spacelike generating the azimuthal direction and $\tilde{\Omega}$ is the orbital period of the helical field $\ell^\mu$. On the other hand, in the co-rotating frame, the Killing field $\ell^\mu\rightarrow t^\mu$, such that $\tilde{V}^\mu\rightarrow\beta^\mu$ (for a discussion on the subtleties associated with this choice, see, e.g., Ref.~\cite{Gourgoulhon:2001ec}). From the perspective of the observer associated with $n^\mu$, the Noether-current, defined in \eqref{eq:Noetherconservation}, decomposes as
\begin{align}
j^\mu= & \ \rho n^\mu+J^\mu, \\
\rho= & -n_\mu j^\mu=i(\bar{\Phi}\eta-\Phi\bar{\eta}),
\end{align}
with local boson number density $\rho$, and spatial current 
\begin{align}
J^\mu=-i(\bar{\Phi}D^\mu\Phi-\Phi D^\mu\bar{\Phi}).
\end{align}
In direct analogy to the rest mass conservation equation for fluids, the global U(1) symmetry of the scalar theory implies the boson number conservation [see \eqref{eq:Noetherconservation}]. With this above decomposition of the current, the conservation law \eqref{eq:Noetherconservation} reduces to
\begin{align}
\mathcal{L}_n \rho=\rho K-\frac{1}{\alpha} D_i(\alpha J^i) .
\label{eq:eulerianconservation}
\end{align}
Correspondingly, the evolution equation for the scalar field---the Klein-Gordon equation---is readily obtained from \eqref{eq:action}:
\begin{align}
\left[\nabla_\mu\nabla^\mu-\partial_{|\Phi|^2}V(|\Phi|)\right]\Phi=0.
\label{eq:KGeq}
\end{align}
Using the foliation defined by $n^\mu$, the Klein-Gordon equation takes the form
\begin{align}
\mathcal{L}_n\eta=\frac{1}{\alpha} D_i(\alpha D^i\Phi)+ K\eta- \Phi\partial_{|\Phi|^2}V(|\Phi|),
\label{eq:EulerianKGeq}
\end{align}
and similarly for the conjugate equation. To proceed, a series of equilibrium conditions, utilizing the helical Killing vector, must be imposed on the matter variables.

Contrary to the matter variables relevant for fluid stars, the scalar matter making up BSs is not time-independent. To understand possible equilibrium conditions, recall that the scalar field ansatz \eqref{eq:scalaransatz}, with $\eta\sim i(\omega-m\Omega/r)\Phi$, contains a harmonic time-dependence of the scalar variables due to the linear time-dependence of the scalar phase. Therefore, here we explore how the ansatz underlying the isolated BS solution, i.e., 
\begin{align}
\mathcal{L}_\ell \Phi=i\tilde{\omega}\Phi,
\label{eq:binaryansatz}
\end{align}
may generalize to a binary system. To that end, we specialize to the co-rotating frame for the remainder of this section. This implies that $\mathcal{L}_\ell\alpha=0$ and $\mathcal{L}_\ell \beta^\mu=0$, which together with \eqref{eq:binaryansatz} imply that $\mathcal{L}_\ell\eta=i\tilde{\omega}\eta$. With these, the Klein-Gordon equation \eqref{eq:EulerianKGeq} reduces to the complex equation
\begin{align}
-\beta^iD_i\eta+i\tilde{\omega}\eta-\alpha K\eta=D_i(\alpha D^i\Phi)-\alpha\Phi\partial_{|\Phi|^2}V(|\Phi|),
\label{eq:fixedfreqeq}
\end{align}
where $\eta=(i\tilde{\omega}\Phi-\beta^iD_i\Phi)/\alpha$ in the co-rotation frame. This second-order elliptic equation for $\Phi$ is directly analogous to the scalar equation obtained from plugging the metric and scalar field ansatz of an isolated stationary BS [given in \eqref{eq:metricansatz} and \eqref{eq:scalaransatz}, respectively] into the Klein-Gordon equation \eqref{eq:KGeq}. Binary BS boundary conditions are $\lim_{|\textbf{x}|\rightarrow\infty}\Phi=0$.

In this approach, the frequency $\tilde{\omega}$ must be chosen. Even in the equal-frequency regime, we expect the binary's frequency $\tilde{\omega}$ to differ from the isolated star's frequency $\omega_1=\omega_2\neq\tilde{\omega}$ due to the increase in binding energy; hence, the naive expectation is that $\tilde{\omega}<\omega_1$. The frequency $\tilde{\omega}$ could be iteratively adjusted to 
achieve a desired property for the binary initial data (e.g. a value for the mass or scalar charge), or a new binary BS solution may be found at fixed $\tilde{\omega}$ (analogous to the case of an isolated BS). For unequal-mass binaries, the binary's frequency is spatially dependent since $\omega_1\neq \omega_2$. A simple choice is to construct a differentiable $\tilde{\omega}$ transitioning from a fraction of $\omega_1$ around the first star to the same (or different) fraction of $\omega_2$ around the second star\footnote{Instead of making an ad-hoc choice for $\tilde{\omega}$, the total charge of the binary may be kept fixed following the approach introduced in Ref.~\cite{Kleihaus:2005me} for isolated BSs. While this approach is convenient for isolated solutions with spatially constant frequency, a generalization to unequal-frequency binary BSs with a spatially dependent frequency may be challenging.}.

The previous ansatz, based on assumption \eqref{eq:binaryansatz}, requires
specifying the binary frequency $\tilde{\omega}$. An alternative
is to specify some profile for the scalar field (or conjugate momentum)
and then assume only the equilibrium
conditions\footnote{Notice, these approaches are manifestly slice-dependent.
Due to the scalar time-dependence, the covariant equilibrium assumptions are
$\mathcal{L}_\ell[ \bar{\Phi}\Phi]=0$, and  $\mathcal{L}_\ell
[\bar{\Phi}\partial_\mu\Phi]=0$. This, in general, implies $\mathcal{L}_\ell
\rho\neq 0$ in the co-rotating frame.}
$\mathcal{L}_\ell|\eta|^2=\mathcal{L}_\ell\rho=0$ [which are implied by
\eqref{eq:binaryansatz}]. Assuming these equilibrium conditions, the two
equations \eqref{eq:eulerianconservation} and \eqref{eq:EulerianKGeq} reduce to
\begin{align}
D_i\left[\alpha (\bar{\Phi}D^i\Phi-\Phi D^i\bar{\Phi}) \right]=-(V^i D_i+\alpha K)(\bar{\Phi}\eta-\Phi\bar{\eta}),
\label{eq:equileq1}
\end{align}
as well as
\begin{align}
\begin{aligned}
& \bar{\eta}D_i(\alpha D^i\Phi)+\eta D_i(\alpha D^i\bar{\Phi})\\
& =-(V^iD_i+2\alpha K)|\eta|^2+\alpha (\bar{\eta}\Phi+\eta\bar{\Phi})\partial_{|\Phi|^2}V(|\Phi|),
\label{eq:equileq2}
\end{aligned}
\end{align}
respectively. Notice, both equations are real. In the parameterization of Sec.~\ref{sec:fixedkineticenergy}, it is natural to interpret these equations as elliptic equations for either $\eta,\bar{\eta}$ or $\Phi,\bar{\Phi}$ (keeping the respective other fixed). Here, no choice of the binary's frequency is required, since the $\tilde{\omega}$-dependence in \eqref{eq:fixedfreqeq} is canceled when adding \eqref{eq:fixedfreqeq} and its complex conjugate to arrive at \eqref{eq:equileq2}. Instead, the implicit assumption is that the free data provides a sufficiently accurate profile for either the conjugate momenta or scalar field to equilibriate the binary by adjusting the other using \eqref{eq:equileq1} and \eqref{eq:equileq2}. 
In principle, either approach could be applied to constructing spinning binary BS initial data, assuming one starts with free data that is sufficiently close to the desired solution.

Ultimately, only the direct implementation of these approaches may test their
applicability in reducing spurious perturbations and equilibriating the matter
in binary configurations, which we leave to future work.

\section{Quality of binary initial data} \label{sec:qualityofinitialdata}

With the formalism in place to compute constraint satisfying binary initial data, we now assess the quality of the constructed data for some specific examples. Note, we are not equilibriating the scalar matter using the ansatz discussed in the previous section. To that end, we compare the physical properties of the free data with those of the constraint-satisfying initial data in Sec.~\ref{sec:simple_super}. We proceed in Sec.~\ref{sec:spurious_osc} by addressing the spurious excitation of oscillation modes insides the BSs of the initial data, which shows up in the emitted GW signal, by using two prescriptions to systematically remove such artifacts. Finally, in Sec.~\ref{sec:ecc_red}, we utilize the standard procedure for reducing orbital eccentricity and discuss its limitations in the context of binary BSs.

\subsection{Superposed free data} \label{sec:simple_super}

\begin{table}[b]
\begin{ruledtabular}
\begin{tabular}{c|c|c|c|c|c|c|c}
Label & Coupling & $m$ & $\omega/\mu$ & $C_i$ & $S_i/M^2_i$ & Setup & $\alpha$\\ \hline
$\mathcal{B}_1$ & $\sigma=0.05$ & $1$ & $0.4$ & $0.12$ & $2.0$ & Axi. & $0$\\ 
$\mathcal{B}_2$ & $\lambda/\mu^2=10^3$ & $0$ & $0.9$ & $0.08$ & $0$ & Axi. & $0$\\
$\mathcal{B}_3$ & $\sigma=0.05$ & $1$ & $0.3, 0.35$ & $0.17,0.14$ & $1.37,1.70$ & 3D & $\frac{\pi}{3}$\\
$\mathcal{B}_4$ & $\lambda/\mu^2=10^3$ & $0$ & $0.86, 0.9$ & $0.12,0.08$ & $0$ & 3D & $0$ \\
$\mathcal{B}_5$ & $\sigma=0.05$ & $0$ & $0.25$ & $0.13$ & $0$ & 3D & $\frac{\pi}{2}$ \\
$\mathcal{B}_6$ & $\sigma=0.05$ & $1$ & $0.3$ & $0.17$ & $1.37$ & Axi. & - 
\end{tabular}
\end{ruledtabular}
\caption{The properties of the isolated constituents of the binaries used throughout the remainder of this work. The configurations with coupling $\lambda$ are solutions in the repulsive scalar model \eqref{eq:repulsivepotential}, while those with coupling $\sigma$ are stars in the scalar theory with the solitonic potential \eqref{eq:solitonicpotential}. Here, $\omega$ is the star's frequency, $m$ is the azimuthal index, $C_i$ is the compactness, and $S_i$ and $M_i$ are its individual spin angular momentum and mass, respectively. Binaries $\mathcal{B}_{1,2,5,6}$ consist of identical stars, whereas $\mathcal{B}_{3,4}$ are made of two stars with different frequencies (and hence, masses, spins, etc.). The mass-ratio of the last mentioned binaries are $q=1.43$ and $q=1.13$, respectively. We also consider non-zero initial complex scalar phase offsets $\alpha$ (as defined at the end of Sec.~\ref{sec:fixedkineticenergy}) between the two stars. Note, in the case of $\mathcal{B}_6$, we vary the phase offset $\alpha\in [0,\pi]$ in Sec.~\ref{sec:headon}; hence, we leave $\alpha$ unspecified here. In the axisymmetric setup, the stars are boosted by the Newtonian free-fall velocity at the given coordinate separation, whereas in the 3D context, the stars are initialized with quasi-circular orbital frequency and spins aligned with the orbital angular momentum; note, however, we also consider a binary with the parameters of $\mathcal{B}_3$ with misaligned spins
in detail in Sec.~\ref{sec:precessingbbs}. }
\label{tab:simple_sup_bbs}
\end{table}

\begin{figure}[t]
\centering
\includegraphics[width=0.48\textwidth]{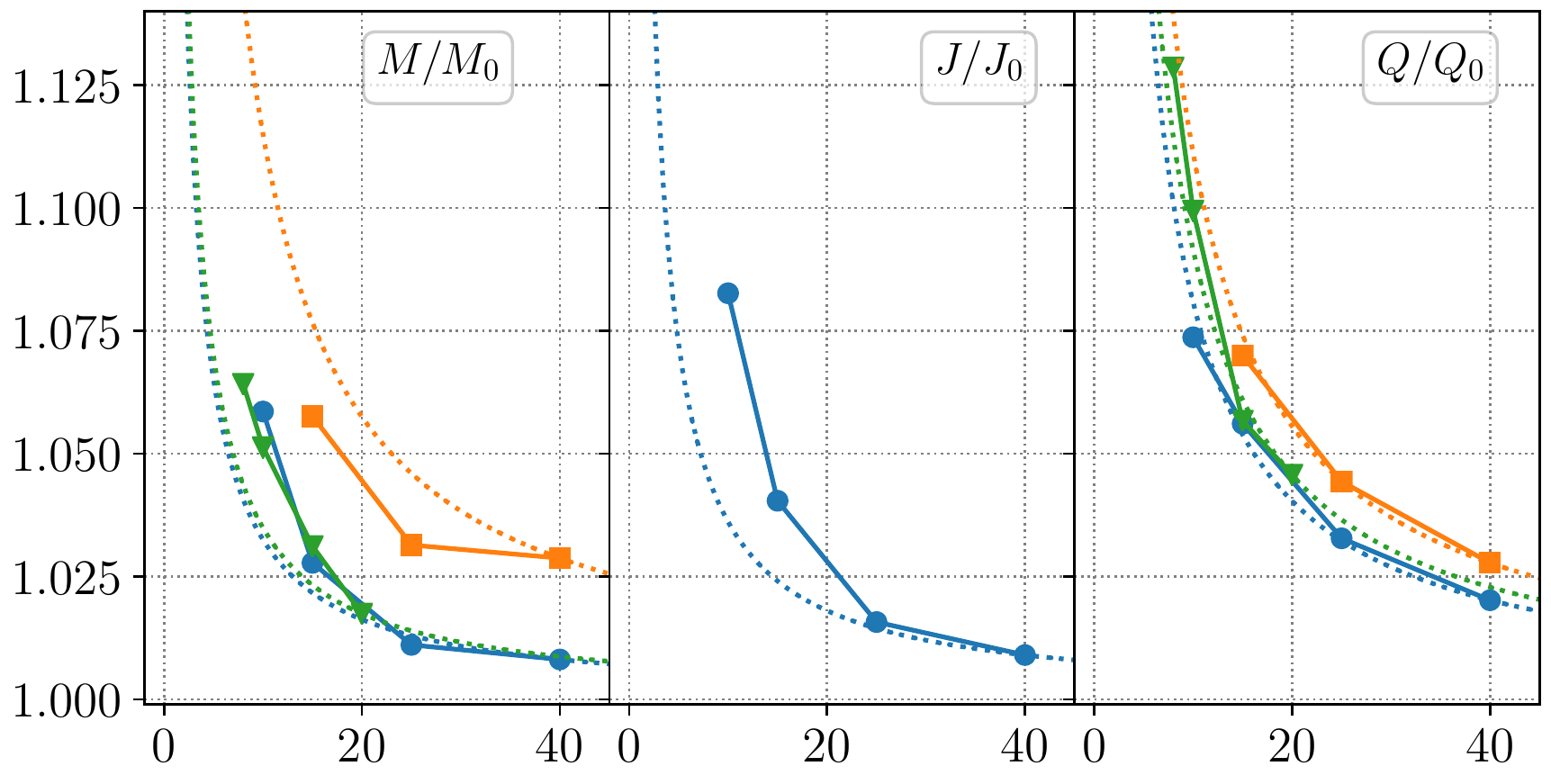}
\includegraphics[width=0.48\textwidth]{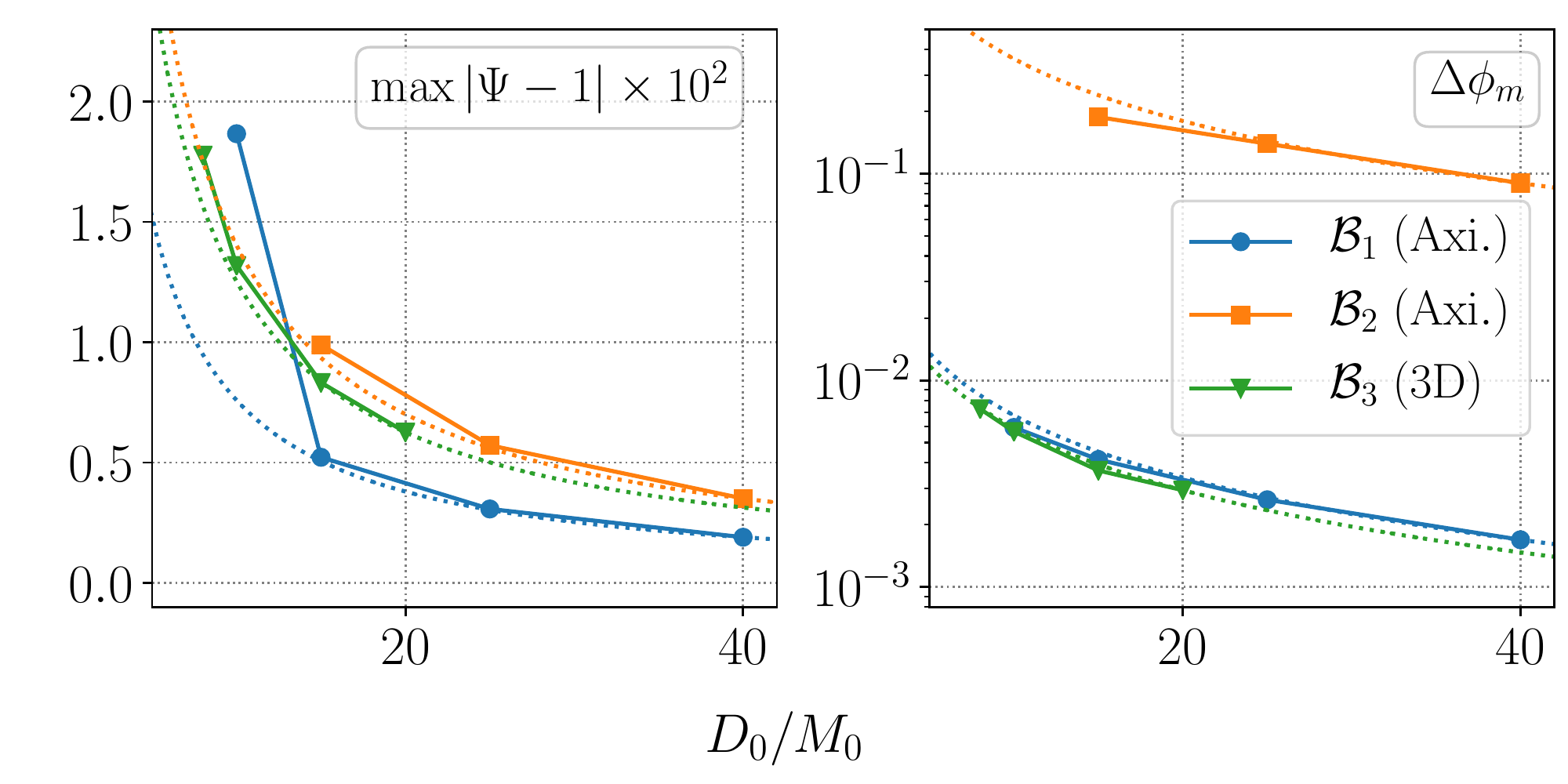}
\caption{The properties of the constraint satisfying binary BS initial data as a function of coordinate separation $D_0$ of the stars. Here, $M$ is the ADM mass, $J$ the angular momentum (defined with respect to the axisymmetric Killing field), and $Q$ the initial charge of the binary configurations with properties summarized in \tablename{ \ref{tab:simple_sup_bbs}}. These are compared with the corresponding quantities at infinite separation of the binary (e.g., $M_0=M_1+M_2$). The amplitude of the spurious oscillations in the stars emerging during the evolution of these binary initial data is defined in \eqref{eq:deltaphi}. Dotted lines indicate the $\propto 1/D_0$ fall-off matched to the point with the largest separations.}
\label{fig:superposeddataproperties}
\end{figure}

To assess the quality of the constructed initial data, we consider a series of binary BS configurations. For now, the focus is entirely on configurations obtained with $f_{(A)}=\hat{f}_{(A)}=1$, i.e., the case of superposed \textit{free} data, as defined in Secs.~\ref{sec:ctsformulation} and~\ref{sec:fixedkineticenergy}. The main properties of the considered configurations are summarized in \tablename{ \ref{tab:simple_sup_bbs}}. For binaries $\mathcal{B}_{1,2,3}$, we analyze the impact of solving the constraint equations on the physical properties of the resulting binary BS initial data as a function of coordinate separation. To that end, in the top panels of \figname{\ref{fig:superposeddataproperties}}, we compare the ADM masses $M$, charges $Q$, and angular momenta $J$ of the constraint-satisfying data to the corresponding quantities of the binary at infinite separation labeled $M_0=M_1+M_2$ and $Q_0=Q_1+Q_2$. The physical properties of the initial data differ from the superposed configuration at infinite separation by up to $\sim 10\%$; these differences roughly scale as $\sim 1/D_0$ with the coordinate separation $D_0$ of the binary. Furthermore, in the bottom left panel of \figname{\ref{fig:superposeddataproperties}}, we show how much the conformal factor $\Psi$ differs from unity. Generally, this difference is at most a few percent for reasonable separations: $\max|\Psi-1|\lesssim 0.02$.

Since the scalar matter is not equilibriated, solving the constraint equations
\eqref{eq:constraints} in this form with superposed free data leads to spurious
oscillations in the constituents of the binary. In the case of binary neutron
stars, these artifacts are identified by monitoring the central density of the
stars during the subsequent evolution of the initial data. Here, we proceed
analogously by tracking the global maximum of the magnitude of the scalar field
$\max |\Phi|$ throughout the first few oscillation periods $T_0$ of excited
modes in the stars. Specifically, we focus on those oscillation modes of
the normalized maximum on each time slice
$\phi_m(t)=\max|\Phi|_t/\max|\Phi|_{t=0}$
and quantify the amplitude of these
perturbations with
\begin{align}
\Delta\phi_m= \frac{\max_{t\in[t_0,t_0+T_0]} \phi_m(t)}{\min_{t\in[t_0,t_0+T_0]} \phi_m(t)} -1.
\label{eq:deltaphi}
\end{align}
In the case of binaries in the repulsive scalar model, measuring $\Delta\phi_m$
with $t_0=0$ suffices, while for binary configurations in the solitonic scalar
theory, we typically extract $\Delta\phi_m$ with $t_0>5T_0$ (once the binary
settles into the dominant oscillation mode).  We find the maximum of the
gauge-dependent U(1)-charge density to track these oscillations equally well.
In the bottom right panel of \figname{\ref{fig:superposeddataproperties}},
we show $\Delta\phi_m$ for different binary configurations and initial
separations. The amplitude of these oscillations increases with decreasing
initial binary separation. At large separations, $\Delta\phi_m\rightarrow 0$,
indicating that the construction of the initial data (without assuming matter
equilibrium) excites these oscillation modes. Notably, the magnitude of
$\Delta\phi_m$ is much smaller in the case of binaries in the solitonic scalar
model, compared with the repulsive scalar model. In the following section, we
analyze spurious oscillations of this kind in more detail, and propose and test
methods to help mitigate these effects.

\subsection{Spurious Oscillations} \label{sec:spurious_osc}

We have seen that the naive choice for the metric and scalar free data, i.e., the superposed free data, leads to potentially significant spurious oscillations in the individual stars in the subsequent evolution of the initial data. To address this issue, it is instructive to consider possible physical mechanisms unique to scalar BSs causing these artifacts. The fundamental feature rendering the fluid star and the BS cases distinct is that the microphysical scales of the latter are \textit{macroscopic}, leading to wave-like phenomena on scales of the star itself. Specifically, the BS can be thought of as composed of a collection of bosons with Compton wavelength $\lambda\sim 1/\mu$ satisfying $M/\lambda\sim CR/\lambda\sim\mathcal{O}(1)$, since $C\sim\mathcal{O}(0.1)$ in the relativistic regimes relevant for this work (see, e.g., \figname{\ref{fig:rvmplot}}). Therefore, there may be distinct processes active in the context of binary BSs affecting the quality of the initial data. 

Self-gravitating solitonic solutions such as BSs consist of a \textit{single}
coherent gravitationally bound state of bosons with energy\footnote{Note, here
and in the following, we use boson ``energy" and ``frequency" interchangeably,
implicitly setting $\hbar=1$.}
$\omega<\mu$. The marginally bound scenario, $\omega=\mu$, separates the bound
states from unbound and \textit{asymptotically} free states with energies
$\omega>\mu$.  Stationary isolated BSs are solutions with bosons of energy
$\omega$ precisely in equilibrium with the gravitational field. However,
perturbations introduced by superposing two stationary BSs and solving the
Hamiltonian and momentum constraints based on such non-equilibriated free
data disrupts this balance. Perturbations may elevate some fraction of the
bound bosons of energy $\omega$ to \textit{(i)} completely free states
(dispersing away from the binary), \textit{(ii)} states that are
gravitationally bound to the binary (as opposed to one of the constituents);
analogous to a wave dark matter halo with solitonic core (see, e.g.,
Ref.~\cite{Hui:2021tkt}), or \textit{(iii)} states gravitationally bound to a
single star, but with energy $\omega$ that is not at equilibrium with the
gravitational sector.  All these processes are, in principle, able to excite
oscillation modes inside the star, as well as contaminate the GW signal at
early times in a numerical evolution. Note, non-linear scalar self-interactions
may also disturb equilibrium configurations. However, since since we mainly
focus on high compact stars with $\omega\ll\mu$, these effects are suppressed
(see, e.g., Ref.~\cite{Siemonsen:2023hko}).

With these possible sources in mind, in the following we explore several
prescriptions for reducing spurious oscillations in binary BSs.  We first
outline methods commonly used in the context of binary black hole and neutron
star initial data and discuss their effectiveness in the BS context, before
introducing and validating several methods specific to BSs.

\subsubsection{Modified superposition} \label{sec:modsuperpos}

\begin{figure*}[t]
\centering
\includegraphics[width=0.48\textwidth]{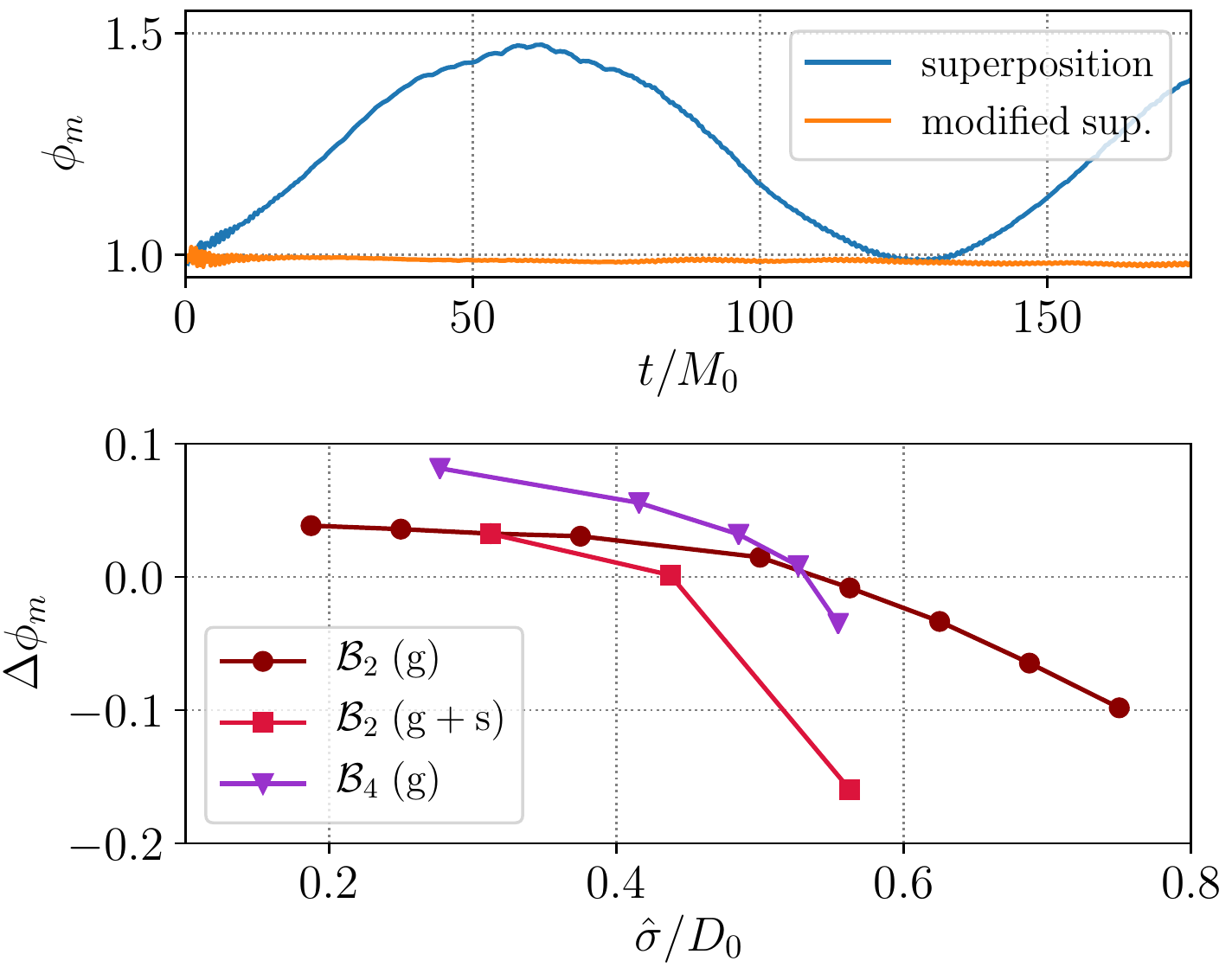}
\includegraphics[width=0.48\textwidth]{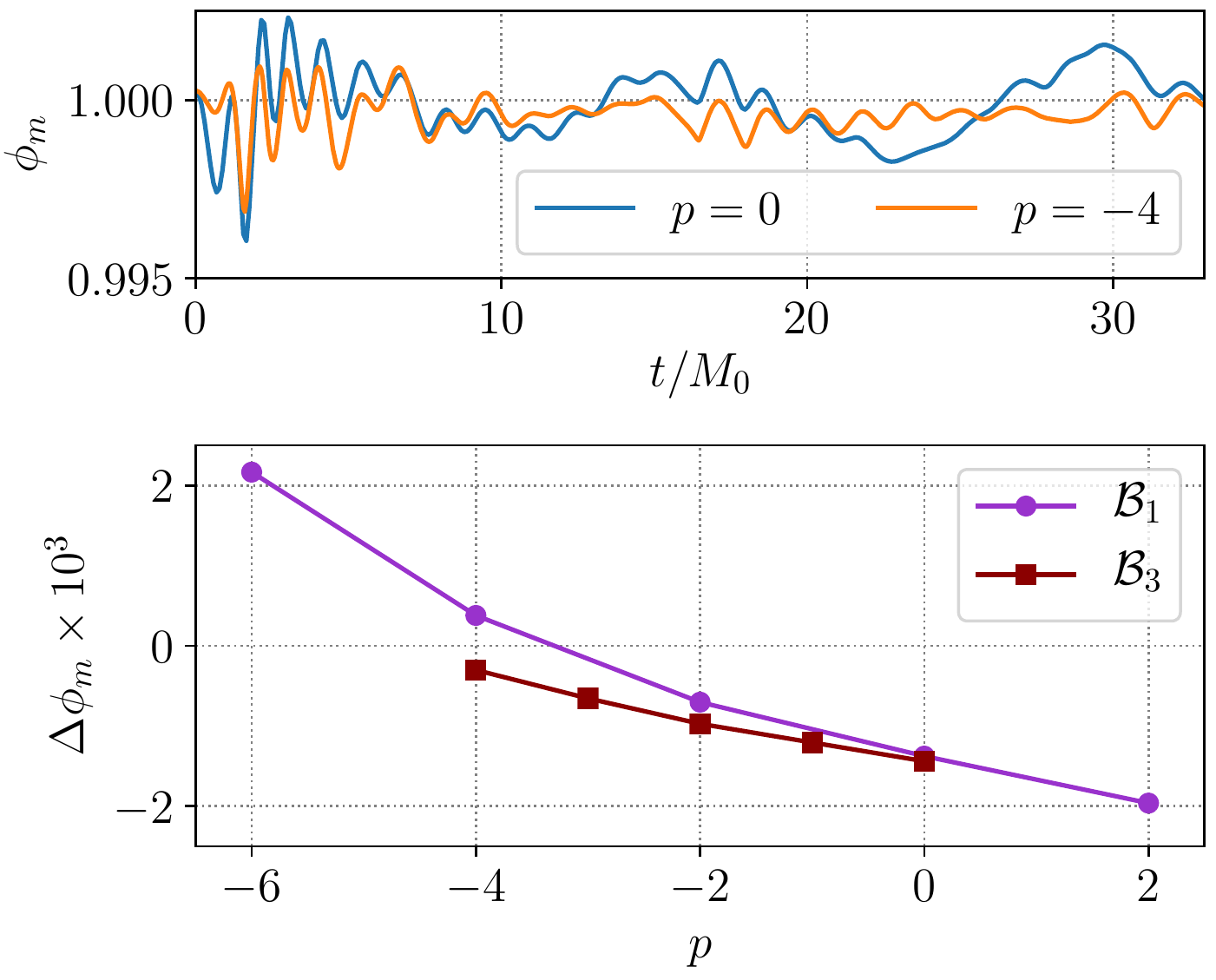}
\caption{We show the temporal evolution of the normalized maximum
on each time slice
$\phi_m=\max|\Phi|/\max|\Phi|_{t=0}$ and amplitude of the perturbations
$\Delta\phi_m$, defined in \eqref{eq:deltaphi}, for selected binary BS
configurations. In all cases shown here, the exponent $\gamma$, defined in
\eqref{eq:att_function}, is set to $\gamma=2$. \textit{(left top)} The behavior
of the maximum $\phi_m$ during the evolution of the binary $\mathcal{B}_4$ (see
\tablename{ \ref{tab:simple_sup_bbs}}) with initial coordinate separation
$D_0=20M_0$, constructed with $f_{(A)}=\hat{f}_{(A)}=1$ (labeled
``superposition") contrasted with the case, where $\sigma_{(A)}/D_0=0.52$ and
$\hat{\sigma}_{(A)}=0$ (labeled ``modified sup."). \textit{(left bottom)} The
amplitude $\Delta\phi_m$ of the spurious oscillations as functions of the
lengthscale $\hat{\sigma}$. 
Cases indicated with ``(g)" correspond to
only metric attenuation,
$\hat{\sigma}_{(A)}=0$ and $\hat{\sigma}=\sigma_{(A)}$, while for those labeled
``(g+s)" both the metric and scalar degrees of freedom are attenuated
$\hat{\sigma}=\hat{\sigma}_{(A)}=\sigma_{(A)}$. The binary $\mathcal{B}_{2}$ has
initial coordinate separation $D_0=40M_0$. \textit{(top right)} The behavior of
the maximum $\phi_m$ during the evolution of the binary $\mathcal{B}_3$ (see
\tablename{ \ref{tab:simple_sup_bbs}}) with initial coordinate separation
$D_0=12M_0$ (constructed with $f_{(A)}=\hat{f}_{(A)}=1$) and rescaling the
conformal kinetic energy in \eqref{eq:conformal_rescale_kinenergy} with $p=0$
as well as $p=-4$. \textit{(bottom right)} We show the amplitude $\Delta\phi_m$
of spurious oscillations emerging during the evolution of binaries
$\mathcal{B}_{1,3}$ with initial coordinate separations $D_0=40M_0$ and
$D_0=12M_0$, respectively. For the latter, we were unable to construct binary
BS initial data with $p<-4$.}
\label{fig:RepulsiveOscillations}
\end{figure*}

We begin by returning to the largest spurious oscillations in the bottom right
panel of \figname{\ref{fig:superposeddataproperties}}, i.e., those in the
head-on collisions of non-rotating stars in the repulsive scalar model. As we
demonstrate below, some of these spurious oscillation artifacts can be removed
by choices of the attenuation functions $f_{(A)}$ and $\hat{f}_{(A)}$,
introduced in Secs.~\ref{sec:ctsformulation} and~\ref{sec:fixedkineticenergy},
respectively. In the case of binary black hole or neutron star initial data, it
is common practice to remove the metric variables of one star at the location
of the other (analogous to approaches introduced in
Refs.~\cite{Marronetti:2000yk,Marronetti:2000rw,Bonning:2003im}), achieved by
non-trivial attenuation functions $f_{(A)}$. 
This can reduce the effect of the superposition on the individual stars.
In order to remove the metric and scalar free data due to one
of the stars at the coordinate location of the other, we choose
\begin{align}
f_{(1)}(\textbf{x})=1-\exp\left[-\frac{|\textbf{x}-\textbf{z}_{(2)}|^\gamma }{\sigma_{(1)}^\gamma }\right],
\label{eq:att_function}
\end{align}
and the corresponding $(1)\leftrightarrow (2)$, as well as scalar attenuation functions $\hat{f}_{(A)}$ with associated length scales $\hat{\sigma}_{(A)}$. Here, $\textbf{z}_{(2)}$ is the initial coordinate position of the second star (as defined in Sec.~\ref{sec:ctsformulation}), whereas the length scales $\sigma_{(A)}$ and $\hat{\sigma}_{(A)}$ set the size of the attenuation region around each of the constituents of the binary. We consider $\gamma=2$ and 4.

We find this approach to be effective in reducing spurious oscillations, as measured by $\Delta\phi_m$, only for BS solutions in the repulsive scalar model \eqref{eq:repulsivepotential}. In fact, applying this approach to stars in the solitonic models \textit{worsens} the artificial oscillations, requiring a different prescription to handle the latter, as described in detail in the next section. In the left panels of \figname{\ref{fig:RepulsiveOscillations}}, we show the time-dependence of the maximum of the scalar field magnitude, as well as the dependence of the amplitude of the spurious oscillations on the lengthscales associated with the attenuation function introduced in Eq.~\eqref{eq:att_function}. As can be seen from the figure, increasing the attenuation length scales (at fixed separation), decreases the amplitude $\Delta\phi_m$. Around $\sigma_{(A)}/D_0\approx 0.5$, the spurious oscillations are minimal, and increase in amplitude for $\sigma_{(A)}/D_0\gtrsim 0.5$. Therefore, we find that in all cases considered, tuning the length scales relevant in \eqref{eq:att_function} results in binary BS initial data with significantly reduced spurious oscillations. Finally, considering $\gamma=4$ (as opposed to $\gamma=2$) results in no qualitative difference to the behavior shown in left panels of \figname{\ref{fig:RepulsiveOscillations}}. 

Besides improving the binary BS initial data by removing spurious oscillations, this modified superposition approach turns out to be \textit{necessary} in the case of highly compact binary inspirals at moderate separations within the repulsive scalar model \eqref{eq:repulsivepotential}. Specifically, we focus on a binary BS with $\omega_1=\omega_2=0.86\mu$, at initial coordinate separation $D_0/M_0
=20$ and quasi-circular boost velocities. For this, we find that superposed free data, i.e., with $f_{(A)}=\hat{f}_{(A)}=1$, results in premature collapse of each individual star to a black hole after $\approx 50M_0$. In contrast, with $\sigma_{(A)}=D_0/2$, the individual stars remain stable throughout the inspiral of length $\sim \mathcal{O}(10^3)M_0$ up to the point of contact. On the other hand, in Ref.~\cite{Siemonsen:2023hko}, we found the simple choice $f_{(A)}=\hat{f}_{(A)}=1$ to be sufficient to successfully evolve the binary with $\omega_1=\omega_2=0.9\mu$ in the same family of solutions; hence, the attenuation is necessary for more relativistic BS solutions. Note, similar premature collapse was observed in Ref.~\cite{Helfer:2018vtq}.

\subsubsection{Conformally rescaled kinetic energy} \label{sec:conformalrescaling}

\begin{figure}[t]
\centering
\includegraphics[width=0.48\textwidth]{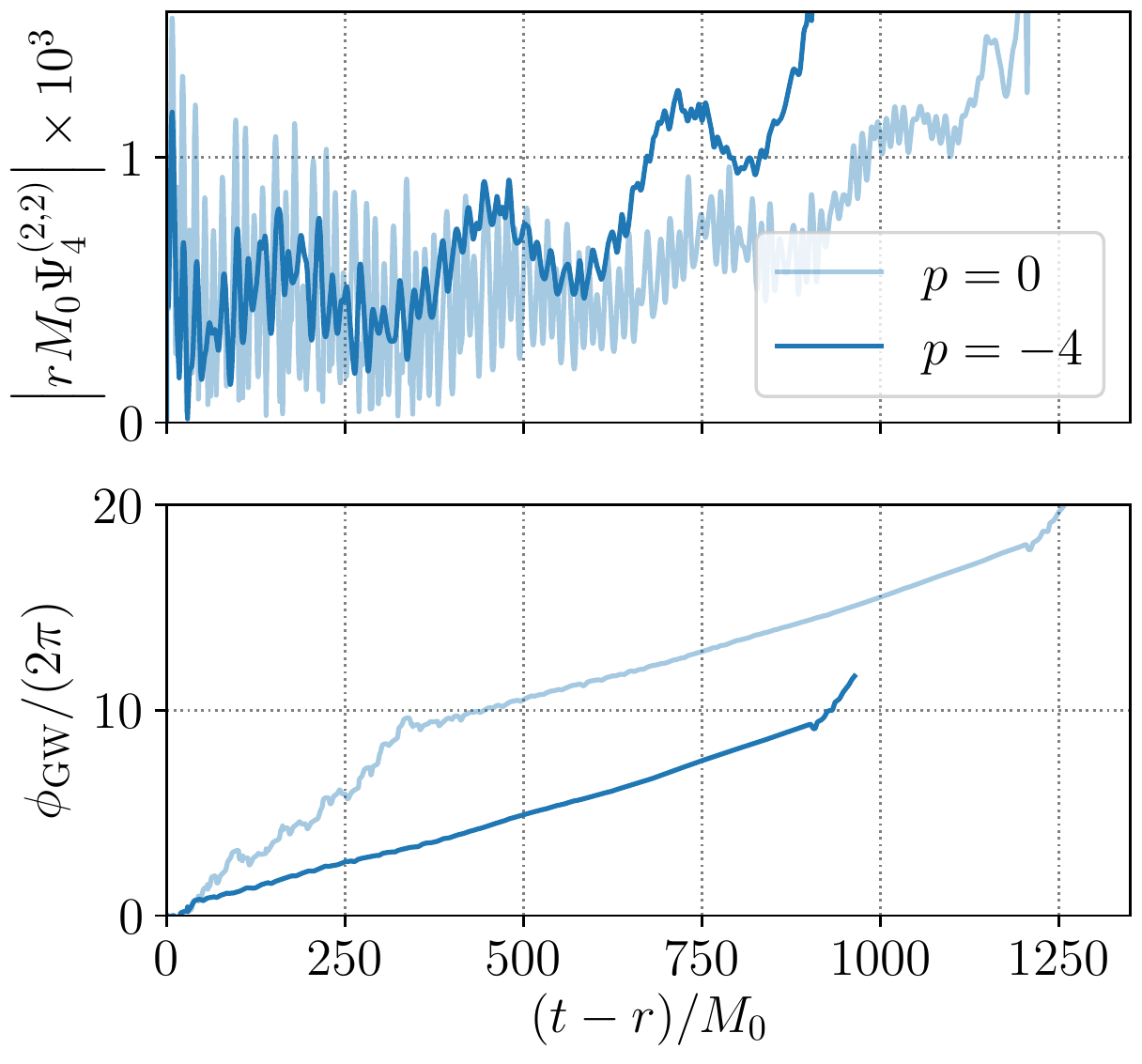}
\caption{The GW amplitude \textit{(top)} and phase \textit{(bottom)} emitted by the (aligned-spin) binary $\mathcal{B}_3$ with initial separation $D_0=12M_0$. The two curves correspond to the evolution of initial data constructed using $p=0$ or $p=-4$ in \eqref{eq:conformal_rescale_kinenergy}. Specifically, the GW phase $\phi_{\rm GW}$ is the complex phase of the Newman-Penrose scalar $\Psi_4$, whereas the GW amplitude is the magnitude of the projection of $\Psi_4$ onto the $(\ell,m)=(2,2)$ mode of $s=-2$ spin-weighted spherical harmonics on a coordinate sphere at radius $r=100M_0$. Note, the binary orbits are eccentric with eccentricity $e\approx 0.1$, resulting in modulations with period $\approx 250M_0$ as is most striking in the top panel (eccentricity reduction is discussed separately in Sec.~\ref{sec:ecc_red}).}
\label{fig:GWcontamination}
\end{figure}

As alluded to above, we find attenuating the metric and scalar free data to only be beneficial in binary BSs in the repulsive scalar theory. For the solitonic case, we return to the notion of energy levels determined by the frequency $\omega$ of the scalar bound state. In the context of the scalar variables, this frequency is set by $\partial_t\Phi$, which enters the kinetic energy $\sim|\eta|^2$ according to \eqref{eq:kineticenergy}. For instance, for isolated BSs $\partial_t\Phi=i\omega\Phi$, and similarly $\eta\sim (\omega-m\Omega/r)\Phi$ (in the coordinates introduced in Sec.~\ref{sec:isolatedstar}). To reduce the impact of the superposition on the BSs, as discussed in Sec.~\ref{sec:spurious_osc}, we modify the kinetic energy of the scalar field\footnote{Note, in principle, one could modify $\partial_t\Phi$ iteratively, even when working with the parameterization of Sec.~\ref{sec:fixedkineticenergy}.}. The kinetic energy combines changes in the frequencies of the individual stars with changes in the local linear and angular momentum due to the orbital motion and the star's intrinsic spin. As such, increasing or decreasing the kinetic energy locally self-consistently within the CTS setup may help address spurious oscillations. To incorporate this in our initial data construction scheme discussed in Sec.~\ref{sec:fixedkineticenergy}, we rescale the physical conjugate moment $\eta$ by an additional power $p$ of the conformal factor:
\begin{align}
\eta=\Psi^{-6-p}\tilde{\eta}.
\label{eq:conformal_rescale_kinenergy}
\end{align}

In the right panel of \figname{\ref{fig:RepulsiveOscillations}}, we
illustrate the impact of the choice \eqref{eq:conformal_rescale_kinenergy} for
different $p$ on the spurious oscillations in the individual stars of two types
of binaries in the solitonic scalar model (in all binary BSs in the repulsive
scalar model, we set $p=0$). In the axisymmetric binary labelled
$\mathcal{B}_1$, we find that the amplitude of the spurious oscillations can be
minimized by adjusting the exponent $p$. In this case, a rescaling
\eqref{eq:conformal_rescale_kinenergy} with $-4<p<-3$ minimizes the spurious
oscillations measured by $\Delta\phi_m$. These oscillations can be addressed
also in the case of the inspiraling binary $\mathcal{B}_3$; however, our
numerical implementation robustly relaxes into a solution to the constraint
equations only for $p\geq -4$. 
The oscillation amplitude is still decreasing with decreasing $p$
for $p=-4$, and this leaves a residual oscillation amplitude roughly a
factor of $5$ smaller compared with the $p=0$ initial data. 
While the spurious oscillation amplitude of all binaries in the solitonic scalar model considered is small, i.e., $|\Delta\phi_m|\sim 10^{-3}$, there is a correlation between reducing these small artifacts and removing a high-frequency contamination from the emitted gravitational waveform.

In \figname{\ref{fig:GWcontamination}}, we show the GW amplitude and phase
extracted from the binary evolution of $\mathcal{B}_3$ constructed using
\eqref{eq:conformal_rescale_kinenergy} with $p=0$. We contrast this with the
signal extracted from the same binary, but with initial data constructed using
$p=-4$. Similar to the amplitude of the spurious oscillations described above,
the high-frequency and large amplitude contamination of GWs at early times changes with
the exponent $p$. As the period of the high-frequency contamination matches the
period of the spurious oscillations in \figname{\ref{fig:RepulsiveOscillations}}, we identify the latter as a source of the GW
contamination. Therefore, the adjustment of the kinetic energy by the conformal
rescaling \eqref{eq:conformal_rescale_kinenergy} aids in reducing GW
contamination. However, as for the spurious oscillations, the trend of the
magnitude of this contamination with decreasing exponents $p<0$ suggests that
if initial data with $p<-4$ could be constructed, the contamination would be
further inhibited. In this case, we find that the amplitude of the
contamination is reduced by roughly a factor of $2$ compared with the canonical
choice $p=0$. Additionally, the evolution of the GW phase shown in \figname{\ref{fig:GWcontamination}} indicates that in the $p=0$ case the amplitude is
dominated by the high-frequency contamination, while in the $p=-4$ case, the
phase evolution is determined primarily by the binary orbit, as opposed to
spurious oscillations. Finally, keeping the superposed free data fixed, but varying
the parameter $p$, results in constraint satisfying initial data with different physical kinetic energy and momentum densities [according to \eqref{eq:conformal_rescale_kinenergy}]. This may lead to different orbital parameters for the binary, explaining the shorter time-to-merger comparing the $p=0$ with the $p=-4$ case in \figname{ \ref{fig:GWcontamination}}.

\subsection{Eccentricity reduction} \label{sec:ecc_red}

The flexibility of our approach to constructing the initial data allows us to, in principle, find constraint satisfying binary BS data resulting in any orbital configuration. Of particular interest in the case of compact binary inspirals are low-eccentricity orbits. Hence, in a last step to improve the quality of our binary BS initial data, we turn to applying common techniques to reduce the eccentricity of compact binary initial data to the binary BSs constructed here. To that end, we first define a notion of the BS coordinate location valid throughout the evolution, we then briefly review the eccentricity reduction methods introduced in Refs.~\cite{Pfeiffer:2007yz,Boyle:2007ft,Mroue:2010re}, and finally, we apply these methods to selected spinning and non-spinning binary BSs.

To define the coordinate positions of the two BSs during a binary evolution, we employ a two-step procedure: first, a rough estimate of the star's position restricted to the initial equatorial plane is obtained by finding the coordinate locations of the local maxima of $|\Phi|$ for spherically symmetric stars, and local minima in the case of rotating BSs associated with each star (i.e., the intersection of the star's vortex lines and the equatorial plane). In a second step, the center of scalar field magnitude within a coordinate sphere $B_{A}$ centered on the previously determined locations of extrema enclosing star $A$,
\begin{align}
z^i_{(A)}(t)=\int_{B_{A}}d^3x |\Phi(t,\textbf{x})|x^i,
\label{eq:centerofmagnitude}
\end{align}
is used as the coordinate position at the given coordinate time $t$\footnote{Note, in the initial timeslice $z^i_{(A)}(0)$ agrees with the coordinate positions $z^i_{(A)}$ defined in Sec.~\ref{sec:ctsformulation} to a large degree; hence, we neglect any potential difference between the two in the following.}. The coordinate separation $d(t)$ of the binary is then simply $d(t)=|\textbf{z}_{(1)}(t)-\textbf{z}_{(2)}(t)|$. Note, the first step is sufficient for stars in the repulsive scalar potential \eqref{eq:repulsivepotential}, as the locations of the extrema are less prone to contamination by spurious oscillations within the stars. In the case of BSs in the solitonic models, however, we find the second step to be crucial particularly for non-spinning stars, as $|\Phi|$ is roughly constant inside each star. In the remainder of this work, we use \eqref{eq:centerofmagnitude} to
define the BS position.

The general procedure to reduce eccentricity, following
Ref.~\cite{Pfeiffer:2007yz}, is to begin with a set of binary initial data,
evolve these for a sufficiently long time to be able to confidently fit for the binary's orbital
parameters, and then compute a correction to the initial radial velocity and orbital frequency to
construct new initial data with lower eccentricity. This process is repeated
until the desired eccentricity is reached. Specifically, we fit for the binary
BS coordinate separation $d(t)$ using 
\begin{align}
\hat{d}(t) = A_{-1}+A_0t+\frac{A_1}{2}t^2+\frac{B}{\omega}\sin(\omega t+\varphi),
\label{eq:doftansatz}
\end{align}
and correspondingly for the binary's radial velocity using the fit
\begin{align}
\hat{v}_r(t) = A_0+A_1t+B\cos(\omega t+\varphi).
\end{align}
Based on this parameterization, the initial orbital frequency $\Omega_0$ and initial radial velocity $v_r$ are corrected by \cite{Pfeiffer:2007yz}
\begin{align}
\begin{aligned}
\Omega_0 & \rightarrow \Omega_0+\frac{B\omega\sin(\varphi)}{2\Omega_0 d_0}, \\
v_r & \rightarrow v_r-B\cos(\varphi),
\end{aligned}
\label{eq:updateformula}
\end{align}
at each eccentricity iteration step, where $d_0=A_{-1}$\footnote{Note, in general $d_0\neq D_0$, as we show below explicitly.}. Working entirely in flat space, in order to translate these orbital parameters to the initial coordinate positions $z^i_{(1)}$, $z^i_{(2)}$ and velocities $v^i_{(1)}$, $v^i_{(2)}$ of the binary constituents, as defined in Sec.~\ref{sec:ctsformulation}, we utilize the Newtonian center-of-mass expressions 
\begin{align}
z^i_{(1)}=r^i  \frac{M_2}{M_0}, \qquad z^i_{(2)}=-r^i \frac{M_1}{M_0}.
\label{eq:centerofmassvar}
\end{align}
Here, $r^i$ is the binary separation with $v^i=\partial_t r^i$, and the 
corresponding velocities $v^i_{(1)}$ and $v^i_{(2)}$ are given by
the time derivatives of the above expressions. We decompose the center
of mass velocity $v^i$ into radial $v_r$ and tangential $v_t$ components as
$v^i=v_r n^i+v_t\lambda^i$, where $n^i=r^i/r$ and $\lambda^i=\partial_t n^i$. The
initial tangential component of the center-of-mass velocity is set by the
initial orbital velocity as well as initial coordinate separation $d_0$ using
$v_t=\Omega_0 d_0$. The constituents velocities are then reconstructed from
$v^i$ utilizing the time derivative of the expressions
\eqref{eq:centerofmassvar}. 
In this context, the orbital eccentricity is defined to be
\begin{align}
e=\frac{B}{\omega d_0}.
\label{eq:eccentricity}
\end{align}
The formulas \eqref{eq:updateformula} are based only on Newtonian
gravity, with radiation reaction and other general relativistic 
corrections assumed to be absorbed into the linear and quadratic
time dependence in \eqref{eq:doftansatz}; 
we note that BSs will also have scalar interactions that
become important during the late inspiral \cite{Siemonsen:2023hko},
particularly for stars with small
compactness and $\omega\approx\mu$, which may introduce extra complications
in performing eccentricity in this way at small separations.

\begin{figure}[t]
\centering
\includegraphics[width=0.48\textwidth]{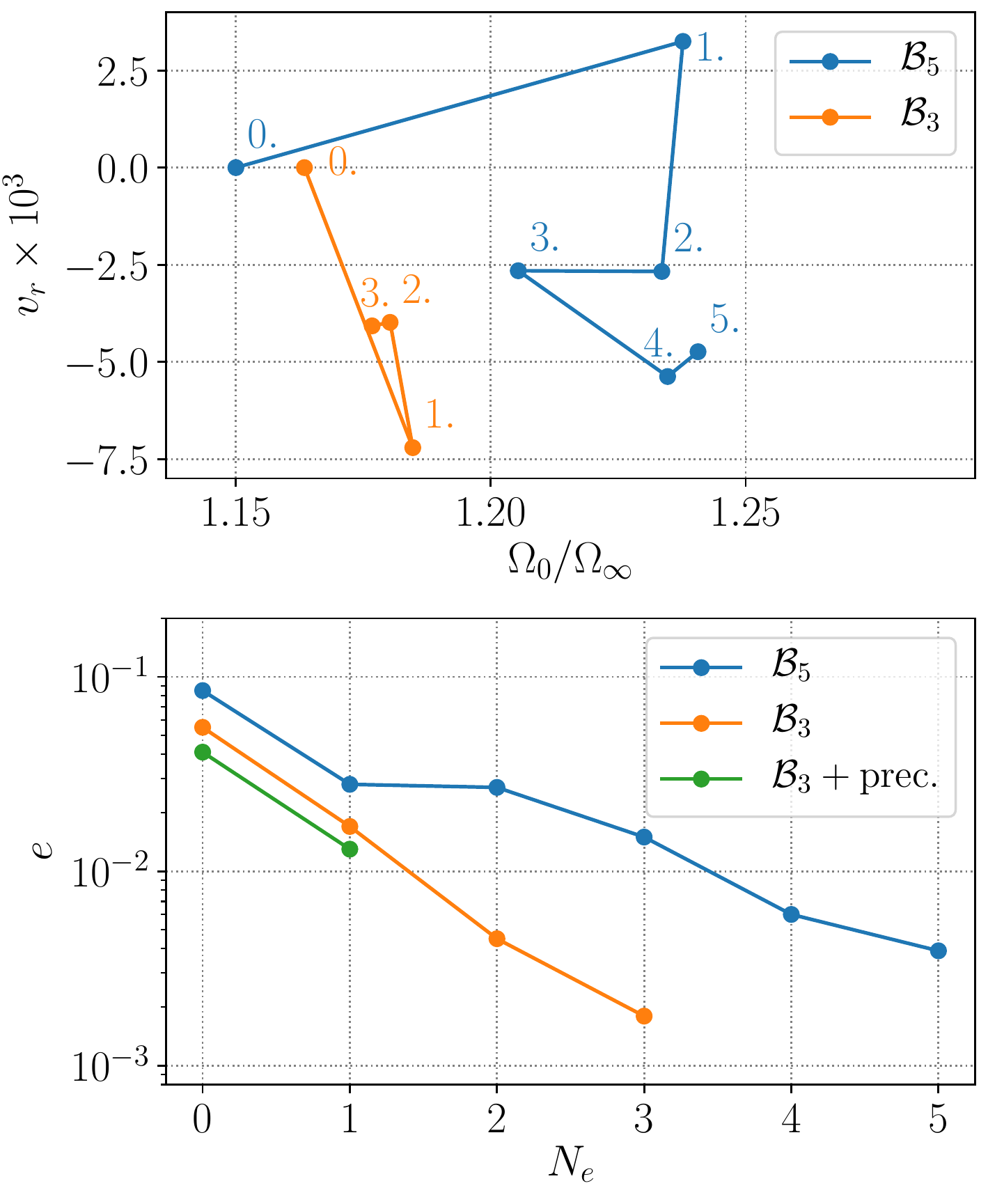}
\caption{\textit{(top panel)} The initial radial velocity component $v_r$ and orbital angular frequency $\Omega_0$ of the binary BS initial data corresponding to $\mathcal{B}_3$ and $\mathcal{B}_5$ (see \tablename{ \ref{tab:simple_sup_bbs}}), throughout the eccentricity reduction procedure, starting from iteration step 0. Here, $\Omega_\infty=(D_0^3/M_0)^{1/2}$. \textit{(bottom panel)} The associated eccentricity, defined in \eqref{eq:eccentricity}, as a function of iteration step $N_e$. We only perform a single iteration step for the $\mathcal{B}_3$ binary with misaligned spins (this case is discussed further in Sec.~\ref{sec:precessingbbs}).}
\label{fig:eccred}
\end{figure}

Applying this machinery to binary BSs, we find that spurious oscillations of the stars and non-equilibriated scalar matter result in high-frequency oscillations of the coordinate separation $d(t)$, limiting the eccentricity reduction. Below eccentricities of $e\sim 10^{-2}$, the fit $\hat{v}_r(t)$ is too uncertain to confidently extract estimates for the subsequent iteration step. Instead, in these cases, we resort to using \eqref{eq:doftansatz} to determine the corrections \eqref{eq:updateformula}. For $e\lesssim 10^{-3}$, the amplitude of the modulation of $d$ due to residual eccentricity reaches the amplitude of the oscillations in $d$ introduced by these spurious oscillations. Therefore, even the fit $\hat{d}(t)$ to the coordinate separation $d$ becomes uncertain, and we terminate eccentricity reduction at $e\gtrsim 10^{-3}$. Furthermore, it is, of course, crucial to minimize oscillations of the stars using the methods discussed in Sec.~\ref{sec:spurious_osc}, i.e., find the exponent $p$ of \eqref{eq:conformal_rescale_kinenergy} and length scales $\{\sigma_{(A)},\hat{\sigma}_{(A)}\}$ in \eqref{eq:att_function}, before attempting to reduce the eccentricity. Especially the choice of \eqref{eq:conformal_rescale_kinenergy} alters the matter's kinetic energy and linear momentum, and hence, the orbital frequency and radial velocities. 

In \figname{\ref{fig:eccred}}, we show the orbital parameters throughout the iterative reduction of the orbital eccentricity for two example binaries. In the case of the spinning and unequal-mass binary $\mathcal{B}_3$, the eccentricity decreases exponentially with the iteration step $N_e$ down to $e\sim 10^{-3}$, where the fitting approach of \eqref{eq:doftansatz} becomes unreliable. In the case of the non-spinning and equal-mass binary $\mathcal{B}_5$, however, the eccentricity decreases only slowly with each iteration step. Consulting the top panel of \figname{\ref{fig:eccred}}, the first iteration step for $\mathcal{B}_5$ resulted in a positive radial velocity. While $v_r$ is a coordinate quantity, and hence, carries limited physical meaning, in binary black hole and neutron star initial data constructions, this is found to consistently satisfy $v_r<0$. Both may be due to fitting to the $d(t)$ time series before all spurious perturbations of the initial data have decayed away sufficiently (e.g., fits of the $N_e=1$ iteration contained only roughly $3/2$ orbits), beyond which we cannot isolate a cause of the slow convergence of $e$ for $\mathcal{B}_5$. Finally, in the case of the precessing $\mathcal{B}_3$ (details can be found in Sec.~\ref{sec:precessingbbs}), we perform only a single iteration step and find similar convergence behavior as for the aligned-spin binary $\mathcal{B}_3$. Note, in precessing cases the spin-interactions (particularly for super-spinning compact objects as considered in $\mathcal{B}_3$) induce physical oscillations of the binary separation beyond residual eccentricity (see, e.g., Ref.~\cite{Buonanno:2010yk}), which we, however, have not observed here. Lastly, in Appendix~\ref{app:centerofmassmotion}, we discuss the linear motion of the center of mass away from the center of the numerical grid, and how to iteratively reduce this artifact, while simultaneously reducing eccentricity.

\section{Binary evolutions} \label{sec:binaryevolutions}

In this section, we illustrate our method for constructing binary BS initial data in the context of head-on collisions and quasi-circular inspirals, including a precessing system. In the case of head-on collisions, we demonstrate explicitly that equal-mass rotating binary BSs exhibit solitonic behavior (resembling the dynamics found in Ref.~\cite{Palenzuela:2006wp}), bouncing off each other when colliding along their spin axes, if the phase offset between the stars is precisely $\pi$. Furthermore, we consider two eccentricity-reduced, quasi-circular inspiraling binary BSs: one non-spinning equal-mass binary, and one super-spinning unequal-mass configuration. We analyze their inspiral dynamics, show that non-trivial scalar interactions result in strong deviations from the dynamics of binary black holes or neutron stars (analogous to what was found in Ref.~\cite{Siemonsen:2023hko}), and characterize the GW strain. Finally, we consider a super-spinning, precessing binary BS inspiral at moderately low eccentricity, analyze the merger dynamics, and show the resulting gravitational waveform. 

To that end, we employ the methods developed above in Sec.~\ref{sec:method}. In particular, we use the source parameterization introduced in Sec.~\ref{sec:fixedkineticenergy} (the scalar matter is not equilibriated with approaches outlined in Sec.~\ref{sec:scalarequilibrium}). Since we focus on BSs in the solitonic scalar model, we set the attenuation coefficients (introduced in Sec.~\ref{sec:modsuperpos}), as well as the conformal rescaling parameter $p$ (introduced in Sec.~\ref{sec:conformalrescaling}), to zero unless otherwise stated.

\subsection{Head-on collisions} \label{sec:headon}

In this section, we explore the merger dynamics of two rotating BSs during a
head-on collision along their respective spin axes, focusing on the
$\sigma=0.05$ solitonic scalar model \eqref{eq:solitonicpotential} and binaries
composed of $m=1$ rotating BSs. It is important to note that, in this setting,
we evolve the binary BSs using a generalized Cartoon method, which explicitly
assumes the scalar fields azimuthal dependence follows $\Phi\sim
e^{im\varphi}$, in addition to an axisymmetric metric (see
Appendix~\ref{app:numerics} for details). As a result, any modes violating this
condition will not appear in the evolution. In particular, this implies
\textit{(i)} any non-axisymmetric instability of the form found in
Ref.~\cite{Sanchis-Gual:2019ljs} is suppressed, and \textit{(ii)} the vortex
structure of the solution on the symmetry axis is conserved throughout the
evolution.

We begin by considering the predictions of the remnant map introduced in
Ref.~\cite{Siemonsen:2023hko} for these head-on collisions.  We do not repeat
the details of the construction of the remnant map here (which are found in
Ref.~\cite{Siemonsen:2023hko}), and simply summarize the key features in the
context of the head-on collision of two $m=1$ BSs. This map assumes U(1)-charge
conservation ($Q_{\rm rem}=Q_1+Q_2$) during the merger of two BSs to predict
the qualitative and quantitative features of the merger remnant. In order to
use the remnant map, one must provide a plausible candidate family of remnants.
Due to our evolution methods, any binary composed of two $m=1$ stars results in
a remnant with $m=1$ vortex along the symmetry axis. Therefore, it is natural
to consider that the remnant of the two rotating BSs is another rotating BS of
the same vortex index (unless the combined charge of the binary surpasses the
maximum charge of the family of $m=1$ BSs in this scalar model, in which case
the remnant is likely a black hole). In our axisymmetric setup, any known
linear instability of the rotating BSs in the $\sigma=0.05$ solitonic scalar
model is suppressed, implying that this condition allows the merger
remnant to be a $m=1$ BS. Hence, we can map all properties of any given $m=1$
binary BS, parameterized by their frequencies $\omega_1$ and $\omega_2$, into
the properties of the resulting $m=1$ BS assuming charge conservation. In
particular, in order to consider the kinematics of the merger---and whether
this favors the formation of a $m=1$ BS remnant---we define the relative mass
difference \cite{Siemonsen:2023hko}
\begin{align}
\mathcal{M}=\frac{M_1+M_2-M_{\rm rem}}{M_1+M_2}.
\label{eq:remnantmass}
\end{align}
Here, $M_1$ and $M_2$ are the masses of the individual stars, while $M_{\rm rem}$ is the mass of the rotating BS with charge $Q_{\rm rot}=Q_1+Q_2$ obtained form the remnant map. If a binary configuration has $\mathcal{M}>0$, then the formation of that $m=1$ BS remnant of mass $M_{\rm rem}$ is energetically favorable.
\begin{figure}[t]
\centering
\includegraphics[width=0.49\textwidth]{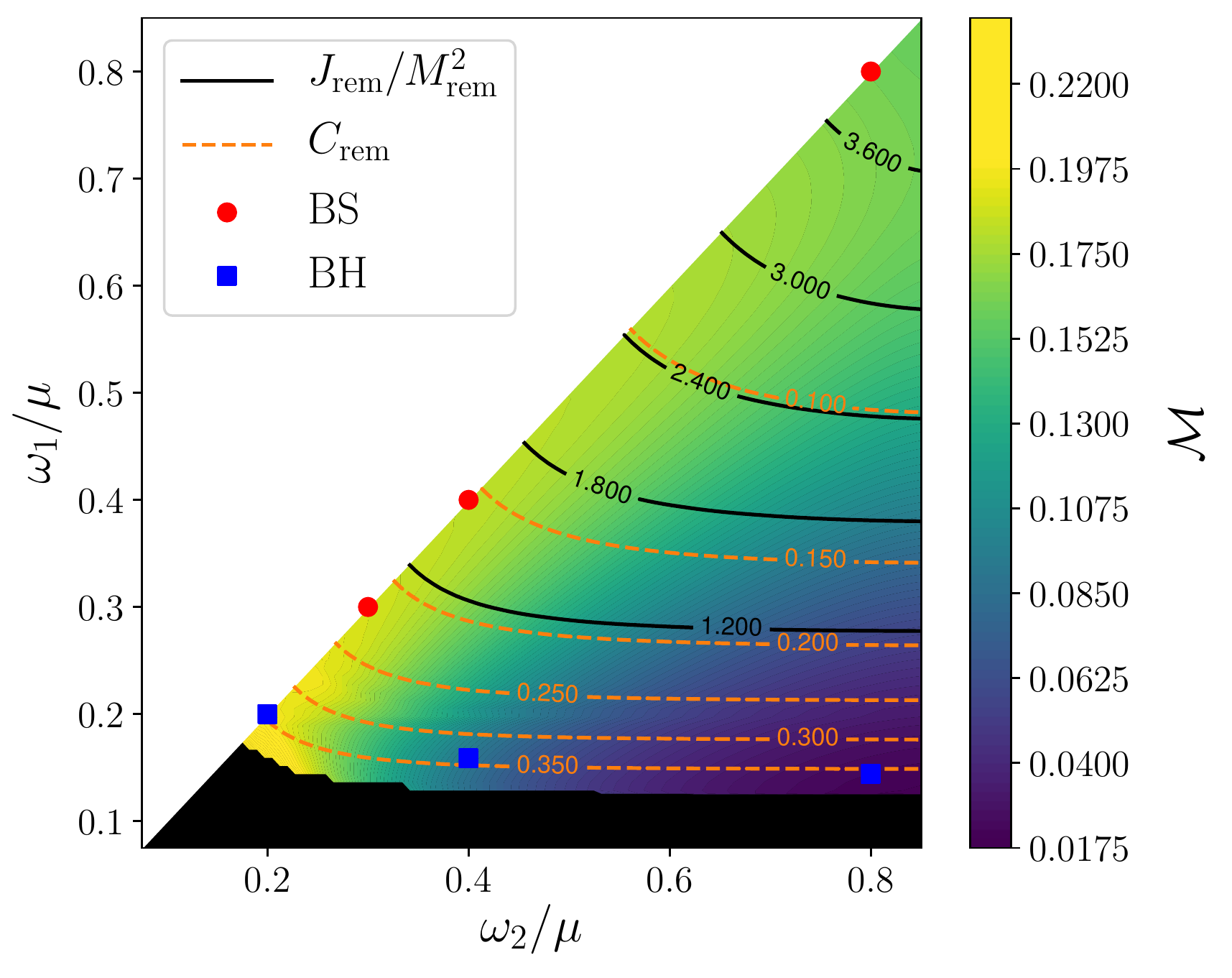}
\caption{The remnant properties of a $m_1=m_2=1$ spinning binary BS head-on
collision (with frequencies $\omega_1$ and $\omega_2$) in the $\sigma=0.05$
solitonic scalar model \eqref{eq:solitonicpotential}, assuming the remnant is a
$m=1$ rotating BS and using the remnant map of Ref.~\cite{Siemonsen:2023hko}
(i.e., assuming U(1)-charge conservation to predict the remnants properties for
each given binary configuration). The dimensionless angular momentum $J_{\rm
rem}/M_{\rm rem}^2$ and the associated remnant compactness $C_{\rm rem}$ are
shown as contours. The mass ratio $\mathcal{M}$ is defined in
\eqref{eq:remnantmass}. We restrict to the $\omega_2>\omega_1$ portion without
loss of generality, and indicate the regions with $Q_{\rm rem}>Q_{\max}$, where
$Q_{\max}$ is the maximum charge of the $m=1$ family of BSs, in black. Finally,
we classify the merger remnants into black holes (``BH") and
$m=1$ rotating BSs (``BS"), and mark the binary that gave rise to the respective
remnant with dots. [Notice, the central (left) ``BS" corresponds to binary
$\mathcal{B}_1$ ($\mathcal{B}_6$), see \tablename{ \ref{tab:simple_sup_bbs}}.]}
\label{fig:remnantmass}
\end{figure}
In \figname{\ref{fig:remnantmass}}, we show the relative mass difference $\mathcal{M}$ across the relevant binary BS parameter space. For all binary configurations shown, it is indeed energetically favorable to form a $m=1$ rotating BS remnant after merger. Hence, the remnant map predicts the merger remnant to be a $m=1$ rotating BS remnant. Note, if the initial phase offset $\alpha$ between the stars is exactly $\alpha=\pi$, then the remnant is not a (parity-even) $m=1$ rotating BS, but rather a double rotating $m=1$ star (i.e., a single, parity-odd rotating star), as we discuss in detail below.

To test this prediction, we perform a
series of numerical evolutions of binary BSs in the axisymmetric setting
described above. The initial data is constructed as discussed in
the previous sections, where for simplicity, the conformal rescaling power in
\eqref{eq:conformal_rescale_kinenergy} is set to $p=0$, and no modification of
the form give in \eqref{eq:att_function} is applied. The initial velocities are chosen
based on the Newtonian free-fall velocity from rest at infinity, and the binary
separation is chosen to be $D_0=10M_0$ initially. Finally, here the phase
offset $\alpha$ (defined at the end of Sec.~\ref{sec:fixedkineticenergy})
between the phases of the rotating stars is set to vanish, $\alpha=0$. (We
consider scenarios varying the value of $\alpha$ below.) For each of the
evolutions, we classify the remnants as either BSs ($m=1$ rotating BSs) or spinning
black holes. In \figname{\ref{fig:remnantmass}}, we show the binary
configurations we numerically evolve, and indicate the
remnant type. First, in the case of equal-mass binaries, i.e., those
with $\omega_1=\omega_2$, the remnant is consistent with the
prediction of the remnant map, except for the case with
$\omega_1=\omega_2=0.2\mu$. In this case, and all other cases indicated as
``BH" in \figname{\ref{fig:remnantmass}}, the merger product collapses to a
black hole during the nonlinear merger process. 
The fact that the threshold for black hole formation is slightly lower
than predicted by the remnant map is likely due to the extra compression
and kinetic energy due to the collision.
This explicitly demonstrates that the final remnant of the
head-on collision of two rotating BSs along their mutual vortex line results in
another rotating BS of the same vortex number (or a black hole if the
individual stars are sufficiently compact).
In Ref.~\cite{Nikolaieva:2022cdm}, a similar analysis was performed in the
Newtonian limit dropping all symmetry assumptions. Since they find that the
central vortex line persists throughout the merger, their results are
consistent with our findings here, and suggest that the latter generalize to 3D
if linear instabilities are absent in both the merging BSs and the remnant BS.

\begin{figure}[t]
\centering
\includegraphics[width=0.49\textwidth]{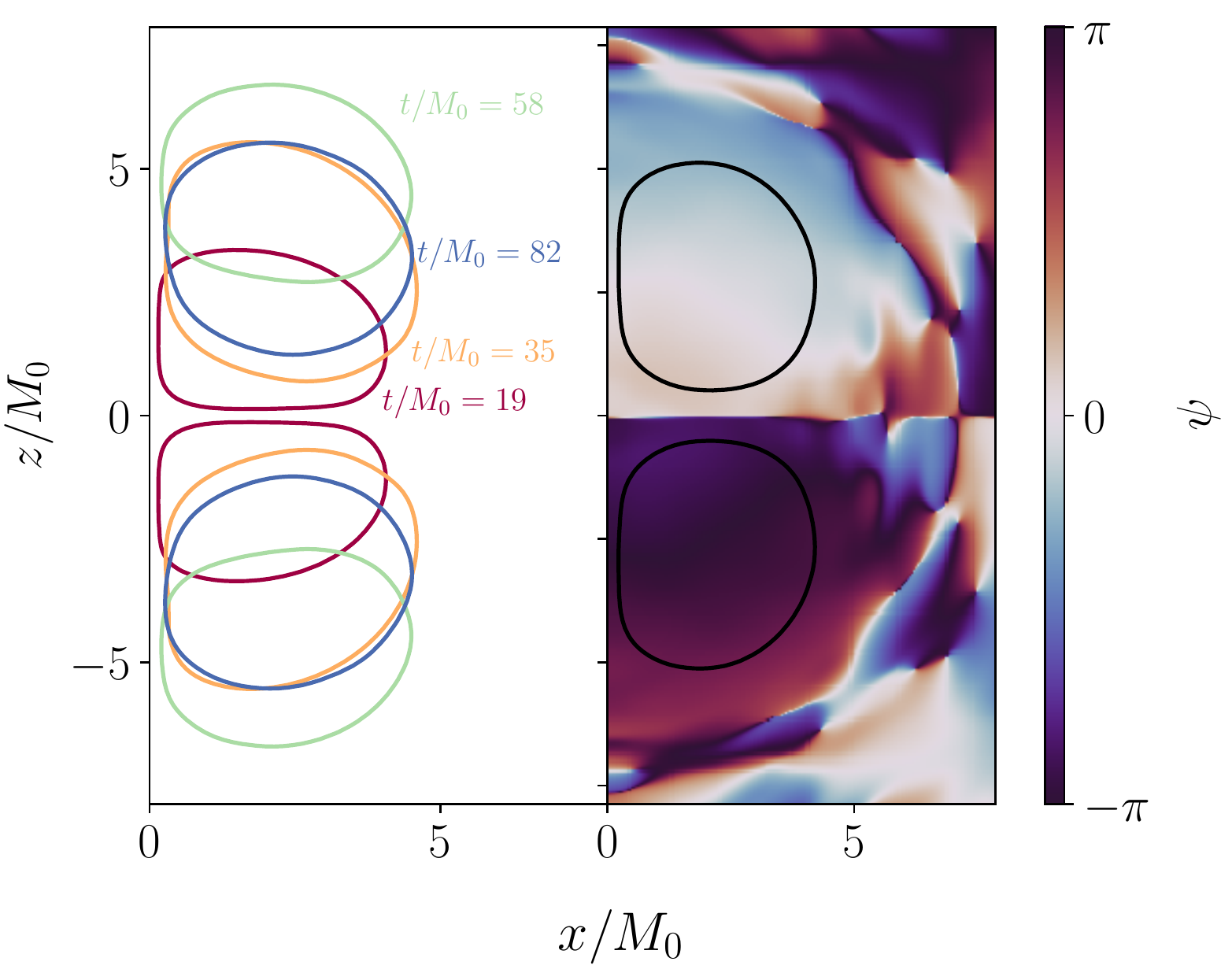}
\caption{We show scalar field quantities on axial slices for the head-on
collision of binary $\mathcal{B}_6$ with initial phase-offset of $\alpha=\pi$.
The $z$-axis is the spin-axis and $z=0$ corresponds to the center of mass of
the system. \textit{(left)} Surfaces of constant scalar field magnitude at
various times during the head-on collision. \textit{(right)} The solution
after it has relaxed at late times ($t/M_0=856$). The black contour line
indicates the surfaces of constant scalar field magnitude, while the color
indicates the phase $\psi$ of the scalar field. This end state resembles the
parity-odd stationary double-BS solutions found in
Refs.~\cite{Kleihaus:2007vk,Gervalle:2022fze}.}
\label{fig:bounce}
\end{figure}

Let us now return to considering head-on collisions of two $m=1$ rotating
binary BS configurations while varying the initial phase offset $\alpha$. We perform
a series of evolutions of the binary configuration $\mathcal{B}_6$ (with
initial separation of $D=10M_0$ and Newtonian free-fall velocities, as before),
with initial phase offsets of $\alpha/\pi\in\{1/4,1/2,3/4,8/9,1\}$. Notice, the
$\alpha=0$ case was found to result in a rotating $m=1$ BS as indicated in
\figname{\ref{fig:remnantmass}}. In \figname{\ref{fig:bounce}}, we show
snapshots of the head-on collision of the binary with maximal phase offset,
i.e., $\alpha=\pi$. As evident there, contrary to the expectation from
the $\alpha=0$ scenario, the two stars bounce off each other upon contact.
After several bounces, the system settles into a state of two spatially
separated scalar distributions with a persistent phase offset of $\pi$, as
shown in the right panel of \figname{\ref{fig:bounce}}. The end state of
this merger is plausibly a stationary solution analogous to those found in
Refs.~\cite{Kleihaus:2007vk,Gervalle:2022fze}. These are parity-odd solutions
resembling two rotating BSs, where gravitational attraction is balanced by
scalar interactions. Physically, this solitonic behavior resembles the dynamics
during the head-on collision of two \textit{non-spinning} BSs with phase offset
$\alpha=\pi$ reported in Ref.~\cite{Palenzuela:2006wp} (and associated
stationary solutions of Ref.~\cite{Yoshida:1997nd}). 

Moving to the cases with $\alpha/\pi\neq1$, we find that the remnant of the corresponding head-on collision is always a single $m=1$ rotating BS at late times. In the case of $\alpha/\pi=8/9$, the system performs a single bounce upon collision, but then merges to a perturbed rotating BS. This demonstrates that, similar to the head-on collision scenario of two non-spinning stars, only the $\alpha=\pi$ configuration exhibits a final state different from a $m=1$ rotating BS. This, of course, limits the validity of the remnant map to those cases with $\alpha/\pi\neq 1$; however, as this is an edge case (similar to two non-spinning stars with phase-offset of $\alpha=\pi$), the impact on the applicability of the remnant map even in these head-on scenarios is minimal.

\subsection{Quasi-circular binaries} \label{sec:quasicirc}

\begin{figure*}[t]
\centering
\includegraphics[width=1\textwidth]{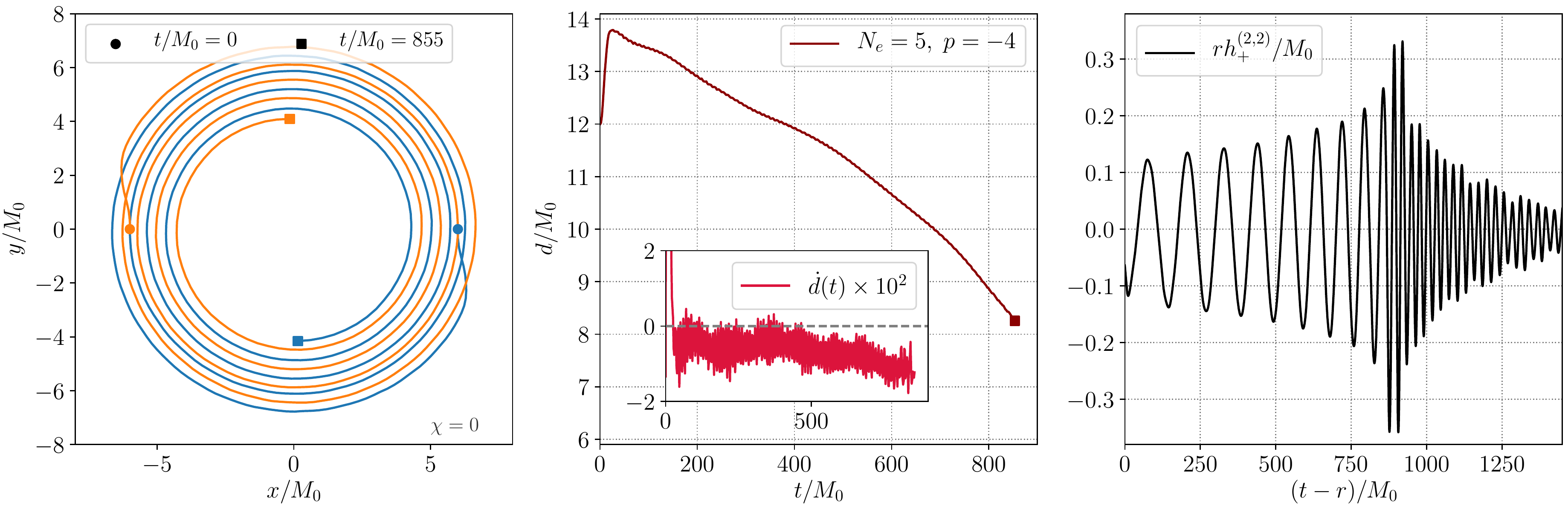}
\includegraphics[width=1\textwidth]{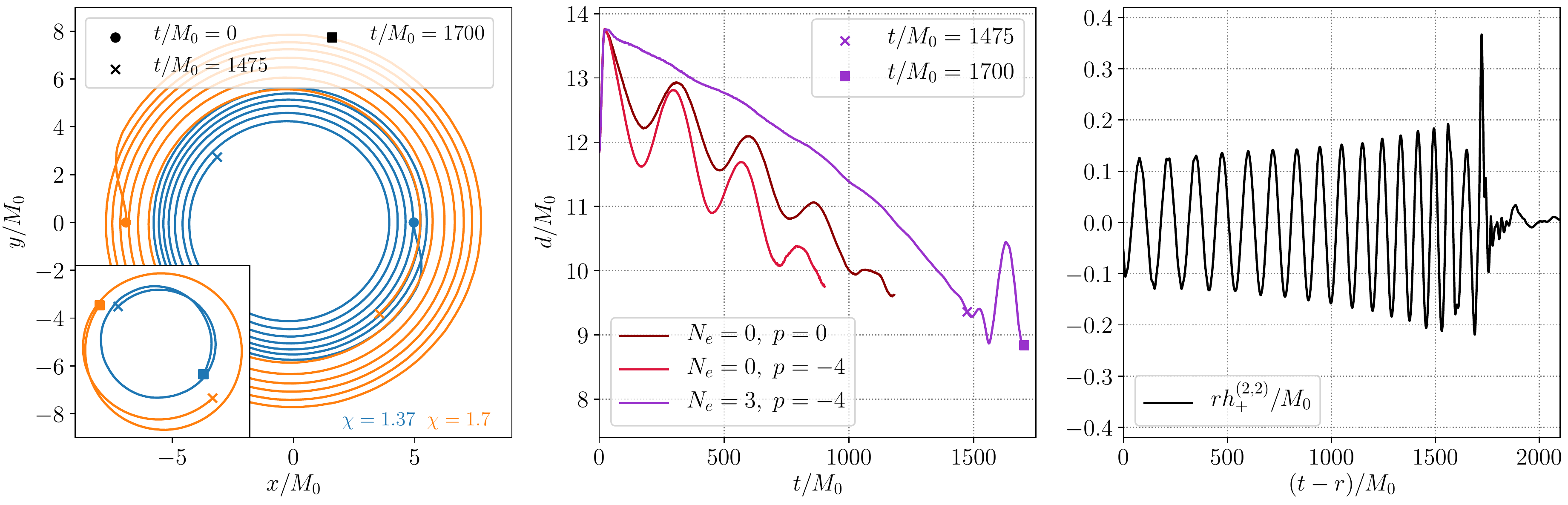}
\caption{
Trajectories and GWs from two quasi-circular binary
BS mergers: an equal-mass, non-spinning binary labelled $\mathcal{B}_5$ \textit{(top row)},
and a unequal-mass aligned spin case $\mathcal{B}_3$ \textit{(bottom row)}.  (The binary
properties are summarized in \tablename{ \ref{tab:simple_sup_bbs}}.) We show
the BS trajectories in the orbital plane [defined according to
\eqref{eq:centerofmagnitude}] throughout the evolution up to the point of
contact \textit{(left panels)}, the binary coordinate separation $d$ as a
function of time \textit{(center panels)}, and the emitted $(\ell,m)=(2,2)$
$s=-2$-weighted spherical harmonic component of the GW strain $h_+^{(2,2)}$
\textit{(right panels)}.  In the left and center panels, we indicate the
initial time (round markers) and the time of contact (square markers). In the
bottom left and center panels, we also indicate the time $t/M_0=1475$
(x-marker) and show the subsequent in-plane star trajectories in the inset in
the bottom left panel up to the point of contact, which is discussed in the
main text. The dimensionless spins, $\chi$, of the binary constituents can be found
in the bottom right corners of the left panels.
The legends in the center panels indicate the eccentricity
reduction step $N_e$ and the rescaling exponent $p$ used in
\eqref{eq:conformal_rescale_kinenergy}.  For comparison, in the bottom-center
panel only, we also show two cases with different values of $p$ before
eccentricity reduction ($N_e=0$).  After eccentricity reduction, $e\approx 4
\times 10^{-3}$ initially for the top row, and $e\approx 1.8 \times 10^{-3}$
for the bottom row.  In the center top panel, we also show the time derivative
of the separation $d$ in the inset.  Notice, the GW strain (right panels)
contains residual high-frequency contamination as discussed in
Sec.~\ref{sec:spurious_osc}; this contamination is shown in detail in
Appendix~\ref{app:contamination}. At the point of contact of the two stars, the
GW frequency is roughly $\omega_cM_0\approx 0.1$ both for 
$\mathcal{B}_5$ and $\mathcal{B}_3$). For the post-merger phase of binary
$\mathcal{B}_5$, we estimate the dominant frequency and exponential decay
timescale of $h_+^{(2,2)}$ (see top right panel) to be $\omega_{\rm
post}M_0\approx 0.23$ and $\tau_{\rm post}/M_0\approx 3\times 10^2$,
respectively. There is a slight drift of the center of mass that is barely
noticeable in bottom left panel corresponding to $v^x_{\rm com}\approx
-1.2\times 10^{-4}$ (all other components are $\lesssim 10^{-5}$; see
Appendix~\ref{app:centerofmassmotion} for a discussion).}
\label{fig:inspirals}
\end{figure*}

We return now to the eccentricity-reduced binary configurations discussed in
Sec.~\ref{sec:ecc_red}, and provide further details on their orbital evolution
as well as GW emission. In particular, we consider the equal-mass, non-spinning
binary $\mathcal{B}_5$, and the aligned-spin super-spinning binary
$\mathcal{B}_3$ of mass-ratio $q=1.43$ with details provided in \tablename{
\ref{tab:simple_sup_bbs}}. In both cases, we focus mainly on the last
eccentricity-reduction iteration step shown in \figname{\ref{fig:eccred}} (i.e., $N_e=5$ and
$N_e=3$ for the non-spinning and spinning binaries, respectively) with
eccentricity $e=4\times 10^{-3}$ in the case of $\mathcal{B}_5$ and
$e=1.8\times 10^{-3}$ in case of the non-precessing binary $\mathcal{B}_3$. In
\figname{\ref{fig:inspirals}}, we show the orbital trajectories, binary
separations and radial velocities, as well as the emitted GW strain for both
configurations.

First, let us focus on the eccentricity reduced binary $\mathcal{B}_5$ (top row
of \figname{\ref{fig:inspirals}}). After initial gauge dynamics (we utilize
damped harmonic gauge and the generalized harmonic formulation, see
Appendix~\ref{app:numerics} for details), the binary settles into a state with
roughly 17\% large coordinate separation of $d/M_0\approx 14$. As can be seen
from the top center panel of \figname{\ref{fig:inspirals}}, the time
derivative $\dot{d}$ of the coordinate separation exhibits large, high-frequency
features throughout the inspiral. These are likely a result of the residual
low-amplitude perturbations of each star; the relative amplitude of the oscillations in this
case is at the $\Delta\phi_m\sim 10^{-3}$ level. Clearly, further eccentricity
reduction cannot rely on a fit to $\dot{d}$, but even a fit to $d$ proves to be
challenging at these eccentricities. Nonetheless, the inspiral dynamics follows
a quasi-circular trajectory up to the point of contact, at a coordinate
separation of roughly twice the star's radii: $d\approx 2R$. As the compactness
of each star is $C=0.13$, we expect qualitative similarities to the inspiral of a binary
neutron star. After the point of contact at $t/M_0=855$, the two
non-spinning stars merge into another \textit{non-spinning} star. As shown in
detail in Ref.~\cite{Siemonsen:2023hko}, the binary $\mathcal{B}_5$ favors the
formation of a $m=1$ rotating remnant BS on purely energetic grounds. However,
it lacks sufficient total angular momentum, and with a phase offset of
$\alpha/\pi=1/2$, is not expected to form a rotating BS remnant, but instead a
non-rotating star, with the residual angular momentum shed in the form of
scalar and gravitational radiation. Hence, after merger, the system rings down
towards a single, non-spinning BS. This is reflected in the GW strain (top right panel of
\figname{\ref{fig:inspirals}}), which exhibits a near
exponential decay in the post-merger phase with a rough decay timescale of
$\tau_{\rm post}/M_0=3\times 10^2$ and a dominant ringdown frequency of
$\omega_{\rm post}M_0=0.23$. It should be noted that, though not obvious from the figure, the GW strain shown in
\figname{\ref{fig:inspirals}} does contain some residual high-frequency GW
contamination of the kind discussed in detail in Sec.~\ref{sec:spurious_osc}.
This is made more apparent in Appendix~\ref{app:contamination}.

We now turn to the inspiral of the binary $\mathcal{B}_3$ shown in the bottom
row of \figname{\ref{fig:inspirals}}, which
exhibits several new features fundamentally different from either
binary black hole or neutrons star coalescences. After the initial gauge dynamics,
the system has a coordinate separation of roughly
$d/M_0\approx 14$. 
Before $t/M_0=1475$ (indicated in the figure with an x-marker),
the binary exhibits a smooth inspiral with decreasing separation.
In the bottom
center panel of \figname{\ref{fig:inspirals}}, we compare the coordinate
distance as functions of time to those cases with no eccentricity
reduction and no conformal kinetic energy rescaling power $p$. Between the
$N_e=0$ cases, the binary with $p=-4$ conformally scaling for the scalar kinetic energy
merges before the otherwise identical binary with $p=0$. After several
iterations of eccentricity reduction ($N_e=3$), the merger occurs later. However,
for $t/M_0\gtrsim 1475$, the coordinate separation $d$ exhibits oscillations,
which culminate in the two stars moving away from each other,
temporarily \textit{increasing} the coordinate separation by approximately
$\sim 10$\% to $d/M_0>10$. After this sudden repulsion, the stars begin to
merge at $t/M_0=1700$ (indicated by a square marker in \figname{\ref{fig:inspirals}}). In the inset of the left bottom panel in \figname{\ref{fig:inspirals}}, we show the in-plane trajectories of both stars from
$t/M_0=1475$ to merger. Qualitatively, the sudden repulsion of the two stars
results in a sudden \textit{increase} of orbital eccentricity for the last
$3/2$ orbits before merger (e.g., the orbital trajectories self-intersect).
This repulsive behavior is likely due to strong scalar interactions 
between the two stars in the late stages of the inspiral. This behavior was
first observed in Ref.~\cite{Siemonsen:2023hko} during the late inspiral and
merger of two BSs in a scalar theory with repulsive scalar self-interactions.
Despite the terminology, systems may exhibit repulsive behavior in both
attractive and repulsive scalar models (see e.g.
Ref.~\cite{Palenzuela:2006wp}). These scalar interactions are exponentially
suppressed by the separation of the binary BS and, hence, only become active in the
late stages of the inspiral, depending on the constituent stars' compactnesses
and frequencies. In contrast to the less relativistic binaries considered in
Ref.~\cite{Siemonsen:2023hko}, the constituents of binary $\mathcal{B}_3$ are
highly compact, with scalar densities that rapidly decay away from the
individual stars. As a result, the scalar interactions increase in importance
over the gravitational interaction only roughly two orbits before the point of
contact. At this stage, however, the strength of the scalar interactions
surpasses that of the gravity, likely resulting in the rapid increase of the
coordinate separation shown in the bottom center and left panels of
\figname{\ref{fig:inspirals}}. This is a feature absent in mergers of
compact binaries composed of black holes and neutron stars, and may serve as a
smoking gun signature to distinguish BS binaries from such cases.

To understand the nonlinear merger dynamics of binary $\mathcal{B}_3$, recall
that the initial phase offset of this binary is $\alpha/\pi=1/3$ and that
$\omega_1\neq\omega_2$ (see \tablename{ \ref{tab:simple_sup_bbs}}). This latter
renders the vortex structure of the binary time-dependent even at the linear
level. Hence, a precise prediction and understanding of the merger outcome
using the remnant map of Ref.~\cite{Siemonsen:2023hko} is challenging. However,
the latter can still be utilized to qualitatively analyze the merger process.
Due to the time-dependent scalar phase, there is no fixed vortex at the center
of mass. Thus, the vortex structure does not prevent the formation of a single
BS, and hence, the final remnant may be a combination of non-spinning stars
including possibly only a single spherically symmetric BS. Both transitions,
two $m=1$ rotating BSs merging to two $m=0$ BSs of the same charge, or a single
$m=0$ of the same charge, are energetically favorable (i.e., the corresponding
quantity analogous to \eqref{eq:remnantmass} satisfies $\mathcal{M}>0$ in the
relevant part of the parameter space). However, since the spins of the
inspiraling super-spinning BSs are aligned with the orbital angular momentum,
the system contains large amounts of angular momentum. While it is plausible
that a single non-spinning BS may shed this angular momentum
sufficiently rapidly during the nonlinear merger (based on findings of e.g.,
Ref.~\cite{Palenzuela:2017kcg}), we find that the final remnant is instead a
\textit{binary} of non-spinning stars, which are flung out away from the center
of mass at high velocities (with the residual angular momentum likely being
converted into orbital angular momentum as the stars come into contact). The
binary separation continues to increase up until we terminate the evolution, at
which point the separation increased to roughly $d/M_0\approx 40$. 
The breakup of the spinning binary at
the point of contact occurs at the locations of the vortices of each spinning
star. As such, the nonlinear merger process of $\mathcal{B}_3$ is qualitatively
similar to what in shown in Fig. 6 of Ref.~\cite{Siemonsen:2023hko}.

\subsection{Precessing binary} \label{sec:precessingbbs}

\begin{figure*}[t]
\flushright
\includegraphics[width=0.961\textwidth]{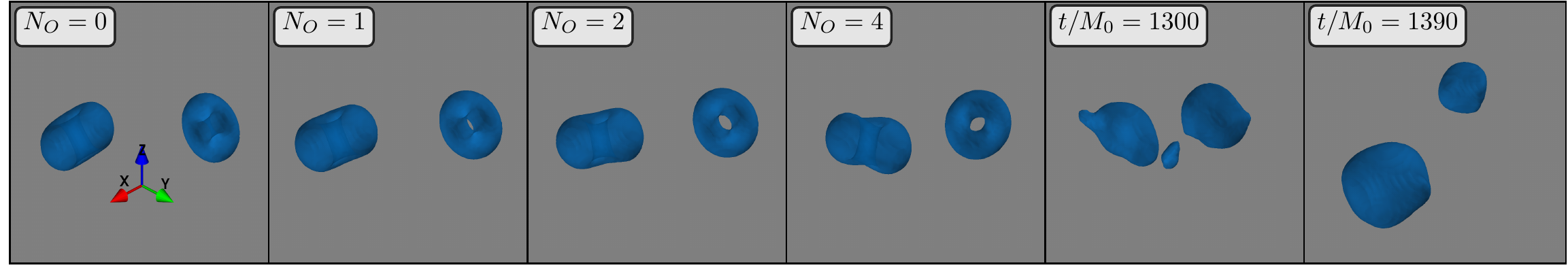}
\includegraphics[width=1\textwidth]{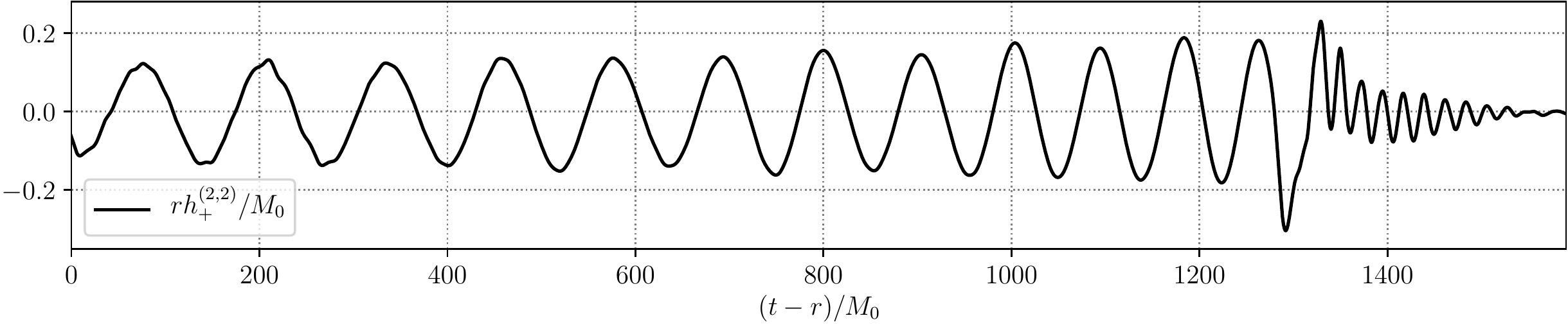}
\caption{
The inspiral and merger of the precessing binary configuration $\mathcal{B}_3$
(see \tablename{ \ref{tab:simple_sup_bbs}}) with initial coordinate separation
$D_0=12M_0$ and initial spin-directions (defined in the main text) so the
spin-vectors point away from the initial boost direction with a
$45^{\circ}$-angle to the initial orbital plane.  We show snapshots with
surfaces of constant scalar field magnitude (3D rendering; \textit{top row})
and the gravitational waveform extracted at $r/M_0=100$ \textit{(bottom row)}.
In the top row, the orientation of the axes (which is the same for all panels)
is shown in the leftmost panel.  The first four panels show the binary during
the inspiral at roughly the same orbital phase after $N_O$ orbits, while in the
last two panels, the merger dynamics is presented roughly at coalescence time
$t/M_0=1300$ and once the resulting non-spinning binary increase its coordinate
separation $t/M_0>1350$.  For this binary, the eccentricity and center-of-mass
drift was reduced in a single iteration step to $e=0.013$ 
and 
$v_{\rm com}\approx 9\times 10^{-5}$.
}
\label{fig:prec_inspirals}
\end{figure*}

Finally, we turn to the precessing binary configuration $\mathcal{B}_3$ (see
\tablename{ \ref{tab:simple_sup_bbs}}) mentioned throughout this work, and
analyze its nonlinear dynamics in detail. The initial data for
binary $\mathcal{B}_3$ is solved using an initial coordinate separation
$D_0=12M_0$ and a conformal rescaling exponent $p=-4$. The spin of each star
is chosen to make a 45$^{\circ}$-angle with the plane containing the initial
positions and velocities of the BSs, such that the component parallel to this
plane is in the opposite direction to the initial tangential velocity.
Recall, the dimensionless
spins of \textit{both} individual stars are above the Kerr-bound $S_i/M^2_i>1$.
As indicated in \figname{\ref{fig:eccred}}, we perform a single
eccentricity and center-of mass velocity reduction step, resulting in an 
eccentricity of $e=0.013$ and linear center of mass drift of $v_{\rm
com}\approx 9\times 10^{-5}$ (with $|v^z_{\rm com}|\approx 7\times 10^{-5}$).

In \figname{\ref{fig:prec_inspirals}}, we show snapshots of the evolution
of these binary BS initial data, as well as the gravitational waveform.
Focusing first on the inspiral, recall that the spin of a rotating BS points
along the vortex line through the center of the star, i.e., perpendicular to
the torus formed by surfaces of constant scalar field magnitude. As can be seen
from the top panel, the binary exhibits rapid precession of the star's spin
vectors throughout the early inspiral. In fact, as the individual stars are
super-spinning, the strength of the spin-orbit and spin-spin interactions 
surpasses that of any binary black hole during the inspiral. As the
eccentricity is relatively large compared to the other
quasi-circular cases, the initial separation is relatively close, and
there is residual high-frequency GW contamination of the type discussed in
Sec.~\ref{sec:spurious_osc}, the typical modulation of the GW amplitude due to
precession cannot be seen in the extracted gravitational waveform. The merger
of this binary is qualitatively similar to the aligned-spin case discussed in
Sec.~\ref{sec:quasicirc}: the two rotating BSs collide to form two non-spinning,
highly perturbed BSs, which move outward from the center of mass. This is
shown in the last two panels of the top row of \figname{\ref{fig:prec_inspirals}}. 
The coordinate separation surpasses $d/M_0>40$ before we terminate the evolution,
and is not clear whether this new binary is gravitationally bound.

\section{Conclusion}

In this work, we tackled the problem of robustly constructing binary BS initial
data satisfying the Hamiltonian and momentum constraints of the Einstein
equations utilizing the CTS formulation.  We analyzed and
tested various approaches to specifying the scalar free data entering these
equations based on superposing isolated boosted BS solutions.  Among these
approaches, we found considering the complex scalar field and its conformally
rescaled kinetic energy as free data to robustly lead to constraint satisfying
initial data that could be readily evolved without further reconstruction
procedures. Beyond a simple superposition of BS solutions, we also reduce the
spurious oscillations induced by non-equilibrium initial data using several
methods.  As suggested in previous studies, we attenuate the superposed free
data of one compact object in the vicinity of the second compact object.  In
addition, here we introduce a new approach where we change the relation between
the initial scalar kinetic energy and the conformally scaled version of this
quantity which is specified as free data.  This reduces the scalar kinetic
energy, and hence, results in less perturbed scalar matter. 
Finally, we successfully reduce the orbital
eccentricity of various mass ratio binary BSs down to the $e\sim 10^{-3}$
level. 

Our procedure for constructing binary BS is highly generic, and thus, is ideally
suited to exploring a vast space of possible binary configurations.
Here we test it for head-on and quasi-circular inspiral
cases, including different mass-ratios, spin magnitudes, and spin orientations.

Ideally, instead of basing the scalar field configuration on superposed BS
solutions, one would solve additional equations imposing a quasi-equilibrium of
the scalar matter with respect to an approximate helical Killing field,
analogous to approaches to construction of equilibriated binary neutron star
initial data. In this work, we briefly discuss several ways in which one might
approach this problem, and some of the complications that may arise, in
particular if one wishes to tackle generic spinning binaries as considered
here. However, we leave an implementation and testing of such an approach to
future work. 

While we were able to efficiently reduce spurious oscillations of the BSs in
the binary, particularly for star solutions in the scalar theory with
repulsive self-interactions, the residual perturbations limit the eccentricity
reduction and contaminate the extracted gravitational waveform. In the cases
we consider with eccentricities $\sim 10^{-3}$, we find further reduction of this
quantity to be challenging as the spurious oscillations in each
star induce high-frequency oscillations of the coordinate separation with
comparable or larger amplitude as the eccentricity. Additionally, even
small-amplitude perturbations in the scalar field making up BSs in scalar
theories with solitonic potential induce large-amplitude high-frequency
contamination in the gravitational radiation at early times during the
numerical evolutions of the initial data. Both artifacts may be suppressed by
solving the CTS constraint equations together with
quasi-equilibrium scalar matter equations. 

With the methods to construct binary BS initial data presented in this work,
accurate waveforms of low-eccentricity non-spinning and (super-)spinning binary
BSs can be obtained. Until the late inspiral, the binary evolution is largely
model-independent, i.e., the inspiral-dynamics is driven by gravitational
effects such as spin-interactions, rather than any mechanism specific to the
scalar matter making up the stars. The resulting waveforms could be used in
to validate and tune inspiral waveform models that would allow for
binary parameters outside the ranges allowed by black holes and neutron 
stars (e.g., Ref.~\cite{LaHaye:2022yxa}).
Likewise, current tests to distinguish binary black
holes and neutron stars from exotic alternatives based on their GW signals
\cite{LIGOScientific:2021sio} could be validated with accurate binary BS
inspiral-merger-ringdown waveforms relying on the initial data constructed in
this work. Another interesting avenue for future work is to study the impact of
relativistic features such as stable light rings and ergoregions of exotic
compact objects on the inspiral gravitational waveform using highly compact
BSs.

\begin{acknowledgments}
We would like to thank Luis Lehner for valuable discussions.  We acknowledge
support from an NSERC Discovery grant. Research at Perimeter Institute is
supported in part by the Government of Canada through the Department of
Innovation, Science and Economic Development Canada and by the Province of
Ontario through the Ministry of Colleges and Universities. This research was
undertaken thanks in part to funding from the Canada First Research Excellence
Fund through the Arthur B. McDonald Canadian Astroparticle Physics Research
Institute. This research was enabled in part by support provided by SciNet
(www.scinethpc.ca), Calcul Québec (www.calculquebec.ca), and the Digital
Research Alliance of Canada (www.alliancecan.ca). Simulations were performed on
the Symmetry cluster at Perimeter Institute, the Niagara cluster at the
University of Toronto, and the Narval cluster at the École de technologie
supérieure in Montreal.
\end{acknowledgments}

\appendix

\section{Numerical setup} \label{app:numerics}

In this appendix, we discuss the details of the 
numerical evolution of the binary BS initial data, which we perform
using the same methods as in Refs.~\cite{Siemonsen:2020hcg,Siemonsen:2023hko}. We evolve the Einstein-Klein-Gordon
system of equations, derived from \eqref{eq:action}, using the generalized
harmonic formulation of the Einstein equations~\cite{Pretorius:2004jg},
including constraint damping terms~\cite{Gundlach:2005eh}. The
spatial fourth-order accurate discretization is achieved using
finite-difference stencils over a compactified grid containing spatial
infinity. At these boundaries, we impose asymptotically flat boundary
conditions both on the metric and the scalar variables. In order to track the
individual stars of a given binary, we utilize adaptive mesh refinement of the
Cartesian grid with a 2:1 refinement ratio (see Ref.~\cite{East:2011aa}). The
time-stepping is achieved using a fourth-order accurate Runge-Kutta
integration. As briefly mentioned in Sec.~\ref{sec:headon}, the axisymmetric
evolutions are performed using a generalized Cartoon method
\cite{Alcubierre:1999ab,Pretorius:2004jg}, which explicitly assumes an
azimuthal Killing field $k^\mu$, such that $\mathcal{L}_kg_{\mu\nu}=0$ and
$\mathcal{L}_k\Phi=im\Phi$. Typically, we require seven mesh refinement levels
with grid spacing $\Delta x/M_0=0.08$ on the finest level when evolving an inspiraling
binary BSs; in the case of axisymmetric evolutions the grid spacing on the
finest level is typically $\Delta x/M_0=0.01$. For all evolutions considered here,
the gauge is specified by setting the source functions $H^\mu=\square x^\mu$ 
according to the damped harmonic
gauge~\cite{Lindblom:2009tu,Choptuik:2009ww}. 

\begin{figure}[t]
\centering
\includegraphics[width=0.485\textwidth]{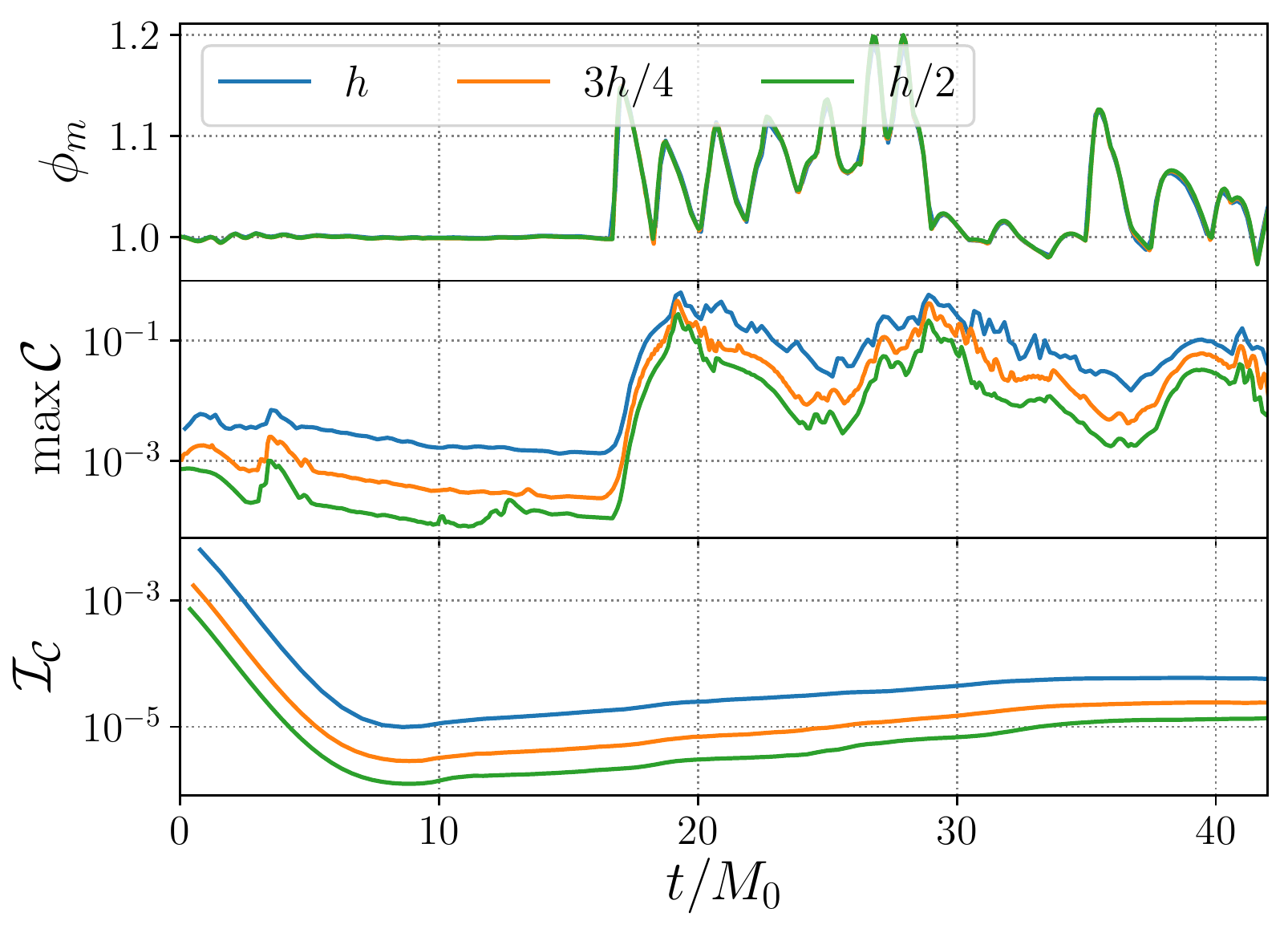}
\caption{
Convergence study of the numerical evolution of the axisymmetric binary
$\mathcal{B}_1$ with properties summarized in \tablename{
\ref{tab:simple_sup_bbs}} (and initial coordinate separation $D=10M_0$), at
three different numerical resolutions, where $h$ is the grid spacing of the
lowest resolution. The point of contact of this head-on collision is roughly at
$t/M_0\approx 16$. The top panel shows $\phi_m=\max|\Phi|/\max|\Phi|_{t=0}$,
the normalized maximum of the scalar field magnitude, while the lower panels
show two different measures of the constraint violation, $\max\mathcal{C}$ and
$\mathcal{I}_\mathcal{C}$ (defined in the text).
The constraint violation is converging to zero at roughly third and fourth order in the middle and bottom panels, respectively.
}
\label{fig:axiconv}
\end{figure}

\begin{figure}[t]
\centering
\includegraphics[width=0.485\textwidth]{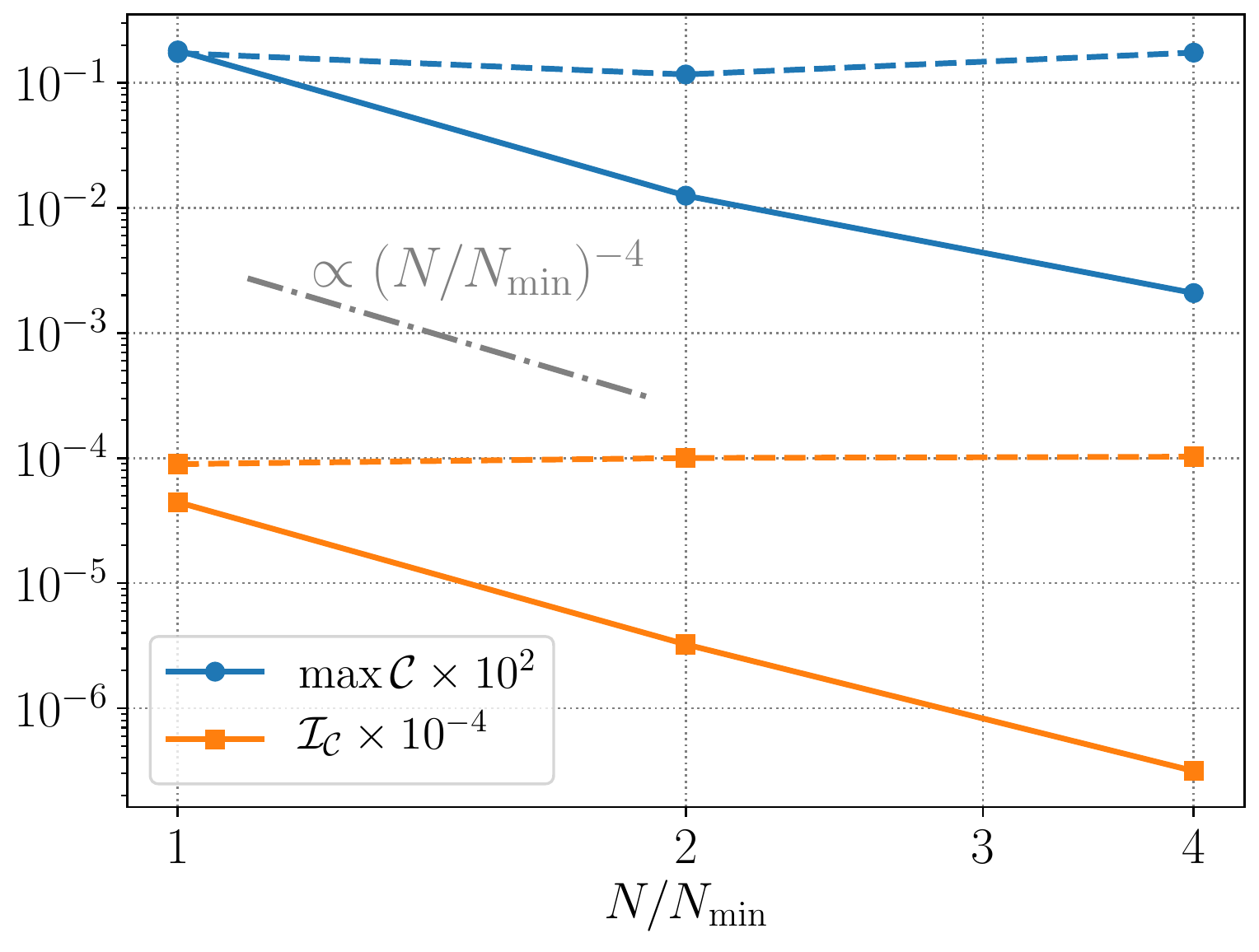}
\caption{The convergence behavior of the two norms, $\max\mathcal{C}$ and
    $\mathcal{I}_\mathcal{C}$ (defined in the main text), of the constraint
    violation of the binary initial data associated with $\mathcal{B}_5$ (see
    \tablename{ \ref{tab:simple_sup_bbs}}) with increasing resolution $N$,
    the number of grid points in each linear dimension, compared with
    the lowest resolution considered $N_{\rm min}$ extracted from subsequent numerical
    evolution at $t/M_0=5$. The solid lines correspond to
    the constraint satisfying initial data constructed with methods developed
    in the main text (with trivial attenuation functions and conformal exponent $p=-4$). 
    The dashed lines indicate the constraint violation
    of the associated free data (i.e., the plain superposition of the two stars).    
    In the case of the constraint satisfying initial data, both norms exhibit (as
    expected) roughly fourth-order accurate convergence towards zero as 
    indicated by the dash-dotted line. The constraint violation of the superposed
    initial data, however, is not converging towards zero, even at very low
    resolutions (note, $N/N_{\min}=1$, or equivalently $\Delta x/M_0=0.16$, corresponds to
    roughly 10 points across each star). }
\label{fig:sbbs_conv}
\end{figure}

\begin{figure}[t]
\centering
\includegraphics[width=0.485\textwidth]{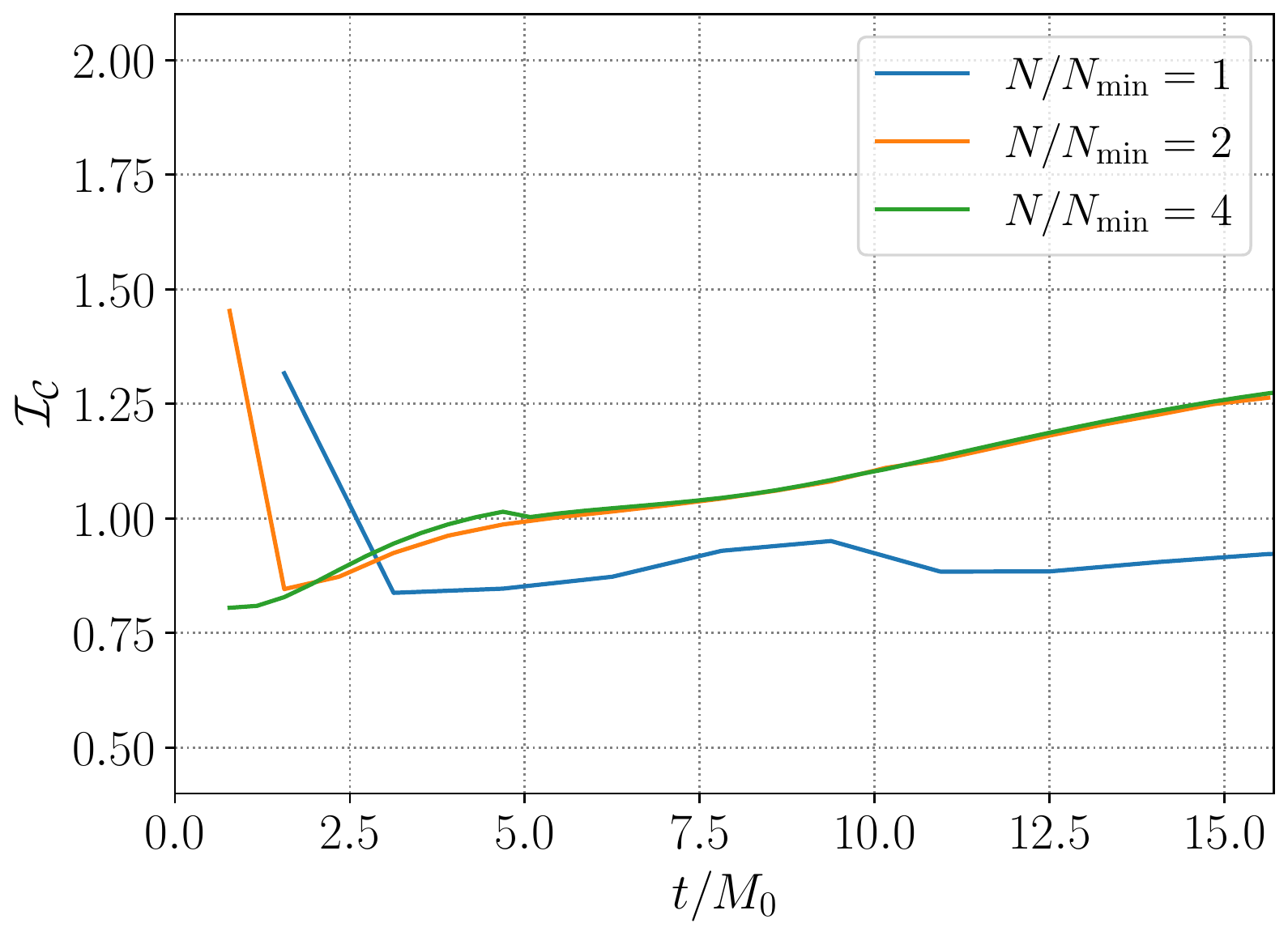}
\caption{The evolution of the integrated constraint violation around the binary $\mathcal{B}_5$ 
starting from plain superposed initial data at three different resolutions $N$ (number of grid
points in each linear dimension) compared to a minimal resolution $N_{\min}$. In \figname{ \ref{fig:sbbs_conv}},
we showed the non-convergence of the constraint violation at early times.}
\label{fig:sbbs_sup_conv}
\end{figure}

We perform convergence tests in order to validate our numerical methods and
quantify the truncation error. Specifically, we measure the constraint
violation given by $\mathcal{C}=M_0\sum_\mu |(H^\mu-\square x^\mu)|/4$ 
throughout the evolutions by computing $\max \mathcal{C}$, as well as
the integrated norm $\mathcal{I}_\mathcal{C}=M_0^{-2}\int d^3x\sqrt{\gamma}
\mathcal{C}$ in a coordinate sphere of radius $20 M_0$. 
In the generalized harmonic formulation, any violation of the Hamiltonian and momentum
constraints on the initial time slice will lead $\mathcal{C}$ to evolve to a non-zero value
(as does truncation error).
In \figname{\ref{fig:axiconv}}, we present the typical convergence behavior of
$\mathcal{I}_\mathcal{C}$ and $\max \mathcal{C}$ for an axisymmetric binary BS
evolution starting from constraint satisfying initial data using the above
numerical evolution setup. In \figname{\ref{fig:sbbs_conv}}, we present a 
convergence study of the constraint violation of the quasi-circular binary initial data
associated with $\mathcal{B}_5$ (see \tablename{ \ref{tab:simple_sup_bbs}}). In
the axisymmetric cases, the maximum of the constraint
violation $\max\mathcal{C}$ is mainly set by the third-order accurate
time-interpolation performed on the mesh refinement boundaries, and as a
result, converges at roughly third order or better in \figname{\ref{fig:axiconv}}. In the three-dimensional evolutions, $\max\mathcal{C}$
is typically reached in the interior of the stars. Hence, in those cases, we
expect a roughly fourth-order convergence behavior, as shown in \figname{\ref{fig:sbbs_conv}}. In both settings, however, the integrated norm
$\mathcal{I}_\mathcal{C}$ converges at approximately fourth
order, as it is less affected by the lower-order interpolation on the mesh
refinement boundaries. 

Also in \figname{ \ref{fig:sbbs_conv}}, we contrast this
convergence with the behavior of the constraint violation with increasing resolution for
initial data composed of the plain superposition of the two stars, i.e. when not solving the constraint equations.
At the lowest resolution, the two are comparable in magnitude. However, the constraint violation
of the free data is \textit{not} converging to zero with increasing resolution.
Resolutions of $N/N_{\min}<1$ are not suitable for evolutions, as 
numerical instabilities associated with unresolved spatial scales (such as the boson mass $\mu$)
appear. Furthermore, we find no evidence that the constraint damping terms we include in
our evolution equations act to damp away the constraint violations from the superposed 
initial data down to a level where they set by truncation error. This is illustrated in \figname{ \ref{fig:sbbs_sup_conv}}. 
Note, the damping timescale associated with the constraint damping is $\tau_{\rm constr.}/M_0=0.5$.

\section{GW contamination} \label{app:contamination}

\begin{figure}[t]
\centering
\includegraphics[width=0.485\textwidth]{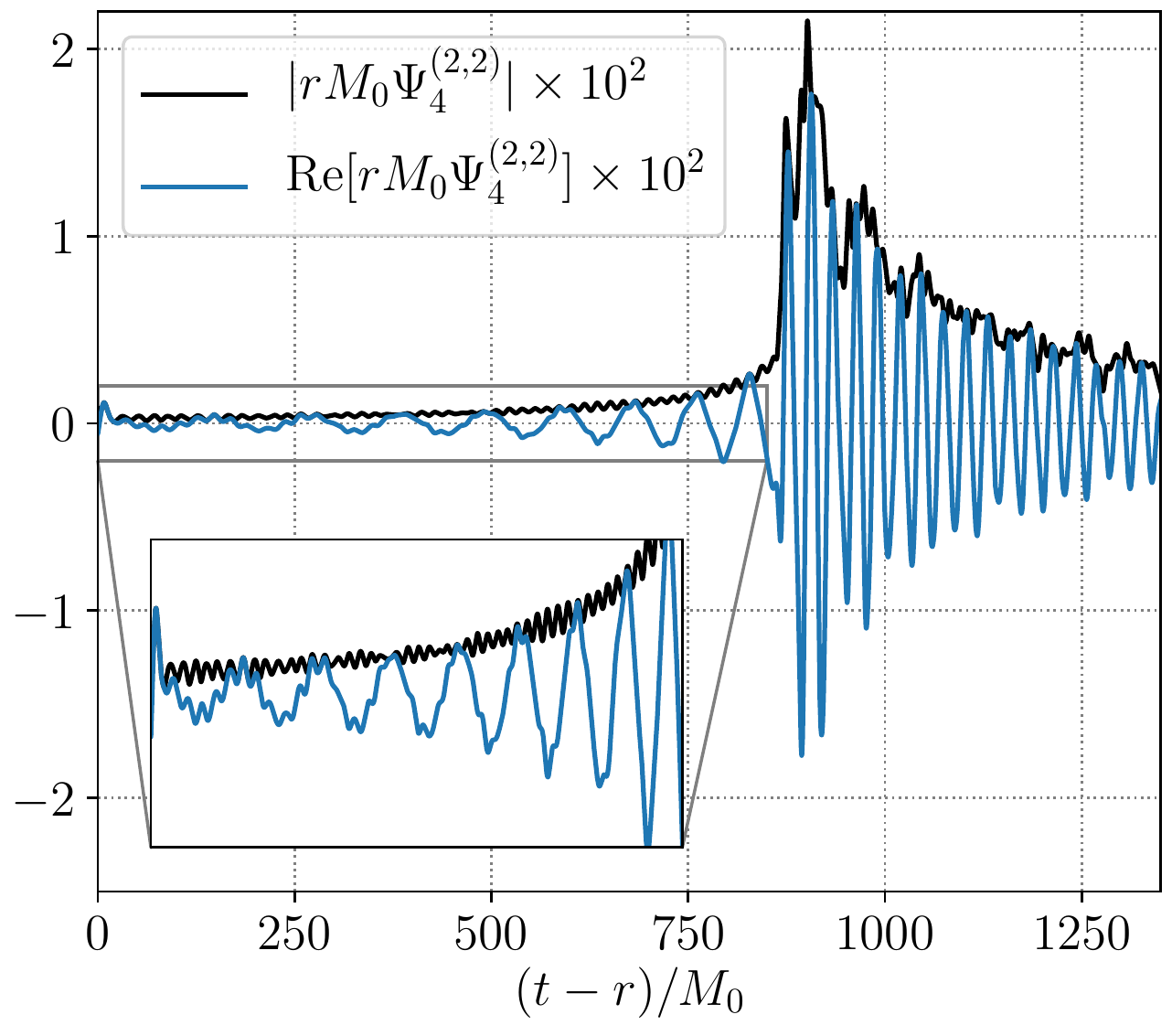}
\includegraphics[width=0.485\textwidth]{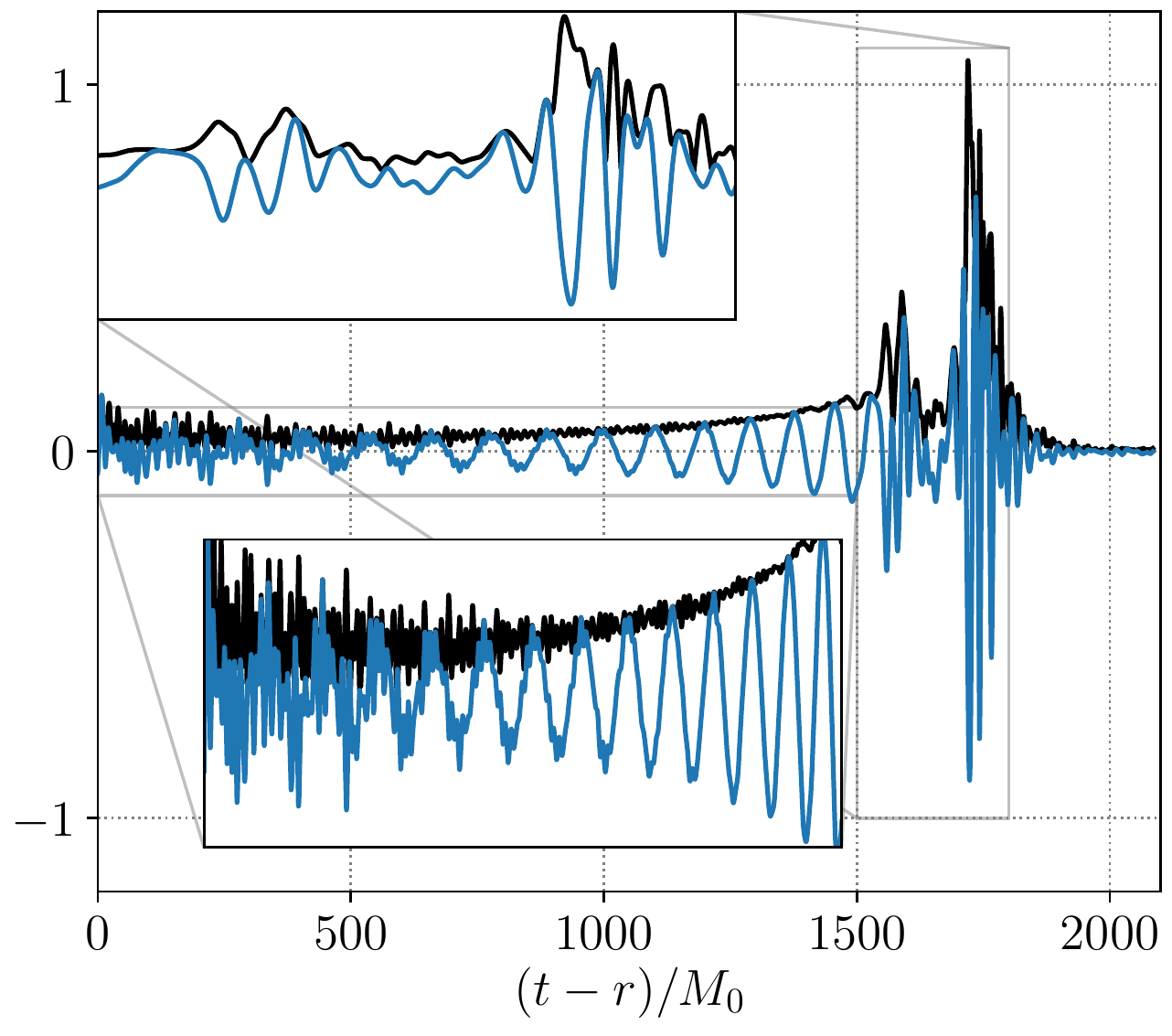}
\caption{We show the $(\ell,m)=(2,2)$ spin-weighted spherical harmonic
components of the Newman-Penrose scalar $\Psi_4$ extracted on a coordinate
sphere of radius $100M_0$. The top panel corresponds to the $N_e=5$ and $p=-4$
binary $\mathcal{B}_5$ shown in the top row of \figname{\ref{fig:inspirals}},
while the bottom panel shows the $N_e=3$ and $p=-4$ binary $\mathcal{B}_3$
shown in the bottom row of \figname{\ref{fig:inspirals}}. This shows the
high-frequency contamination of the gravitational waveform from the binaries at
early times due to residual spurious oscillations and unbound scalar matter in
and around the constituents of the binaries.}
\label{fig:psi4contamination}
\end{figure}

In this appendix, we briefly return to the high-frequency contamination in the
GW emission from the binaries presented in Sec.~\ref{sec:quasicirc}. This
contamination emerges from the residual perturbations present in the
eccentricity reduced binaries $\mathcal{B}_3$ and $\mathcal{B}_5$ constructed
with conformally rescaled kinetic energy using
\eqref{eq:conformal_rescale_kinenergy} with $p=-4$. In \figname{\ref{fig:psi4contamination}}, we present the $(\ell,m)=(2,2)$ spherical
harmonic component of the Newman-Penrose scalar $\Psi_4$, which can be compared
to the corresponding GW strain shown in the right panels of \figname{\ref{fig:inspirals}}. Since $\Psi_4$ is related to the strain by two time
derivatives, it accentuates the high-frequency component from the perturbed
BSs.  As shown in Sec.~\ref{sec:spurious_osc}, this contamination is reduced by
means of the rescaling \eqref{eq:conformal_rescale_kinenergy} of the conformal
kinetic energy. However, with the lowest exponent with which the elliptic
solver was able to find a solution, $p=-4$, some oscillations remain in the
stars, and lead to the high frequency component to $\Psi_4$ evident at early
times in \figname{\ref{fig:psi4contamination}}.

\section{Center-of-mass motion} \label{app:centerofmassmotion}

Within our approach, the initial linear momentum is set to zero using Newtonian
expressions for the initial boost velocities (as discussed in
Sec.~\ref{sec:ecc_red}). We find this to be sufficient for non-spinning
binaries, i.e, the center-of-mass velocity throughout the evolution of the
initial data remains below $v_{\rm com}<10^{-7}$. For highly-spinning quasi-circular binary
initial data, the center of mass of the system exhibits larger
drifts with constant velocity away from origin of the numerical grid. For
the aligned-spin binary $\mathcal{B}_3$, the magnitude of the in-orbital-plane coordinate
velocity of the center of mass is $v_{\rm com}=6\times 10^{-3}$ (the
out-of-plane component is $< 10^{-10}$). Leading post-Newtonian corrections to
the center-of-mass and center-of-momentum velocities, used to initialize the
binaries in this work, are roughly an order of magnitude too small to account
for $v_{\rm com}$. 
This strong drift may be the result of spurious gravitational and scalar radiation emitted
during the first few light crossing times of the binary, as well as the
large spins of the super-spinning binary $\mathcal{B}_3$. 
We address this by measuring the in-plane components $v_{\rm
com}^i$ of the center-of-mass coordinate velocity and subtracting this from the
initial binary velocities $v_{(A)}^i$ of the free data as defined in
Sec.~\ref{sec:ctsformulation}. This is done in tandem with the eccentricity
reduction. Hence, the linear momentum of the binary can be iteratively reduced 
in this way. After two iteration steps, the velocity is reduced by more than an order
of magnitude to $v_{\rm com}=1\times 10^{-4}$ for the binary $\mathcal{B}_3$
with aligned spins. In the case of the precessing binary $\mathcal{B}_3$,
discussed in Sec.~\ref{sec:precessingbbs}, the center-of-mass motion in the direction of the
orbital angular momentum dominates; this we treat iteratively in precisely the
same manner as the in-plane center-of-mass velocity.

\bibliography{bib.bib,bib2.bib}

\end{document}